\documentclass[aps,prc,preprint,amsmath,amssymb,showpacs,superscriptaddress]{revtex4}

\usepackage{graphicx}% Include figure files
\usepackage{dcolumn}% Align table columns on decimal point
\usepackage{bm}% bold math
\usepackage{longtable}
\usepackage{color}
\usepackage{CJK}

\begin{document}

\title{Progress on tilted axis cranking covariant density functional theory for nuclear magnetic and antimagnetic rotation}
\author{J. Meng}%
\affiliation{State Key Laboratory of Nuclear Physics and Technology, School of Physics, Peking University, Beijing 100871, China}
\affiliation{School of Physics and Nuclear Energy Engineering, Beihang University, Beijing 100191, China}
\affiliation{Department of Physics, University of Stellenbosch, Stellenbosch, South Africa}
\author{J. Peng}%
\affiliation{Department of Physics, Beijing Normal University, Beijing 100875, China}
\author{S. Q. Zhang}%
\affiliation{State Key Laboratory of Nuclear Physics and Technology, School of Physics, Peking University, Beijing 100871, China}
\author{P. W. Zhao}%
\affiliation{State Key Laboratory of Nuclear Physics and Technology, School of Physics, Peking University, Beijing 100871, China}
\date{\today}
\begin{abstract}
Magnetic rotation and antimagnetic rotation are exotic rotational phenomena observed in weakly deformed or near-spherical nuclei, which are respectively interpreted in terms of the shears mechanism and two shearslike mechanism. Since their observations, magnetic rotation and antimagnetic rotation phenomena have been mainly investigated in the framework of tilted axis cranking based on the pairing plus quadrupole model.
For the last decades, the covariant density functional theory and its extension
have been proved to be successful in describing series of nuclear
ground-states and excited states properties, including the
binding energies, radii, single-particle spectra,
resonance states, halo phenomena, magnetic moments,
magnetic rotation, low-lying excitations, shape phase transitions,
collective rotation and vibrations, etc. This review will mainly focus on the
tilted axis cranking covariant density functional theory
and its application for the magnetic rotation and antimagnetic rotation phenomena.
\end{abstract}

% 21.60.Jz Nuclear Density Functional Theory and extensions (
%includes Hartree-Fock and random-phase approximations)
%21.10.-k Properties of nuclei; nuclear energy levels
%21.10.Re Collective levels
%21.60.Ev Collective models
%23.20.-g Electromagnetic transitions
%23.20.Js Multipole matrix element
%27.60.+j 90 A 149
%27.50.+e 59  A  89

\maketitle

\section{Introduction}

\subsection{Basic concept}

The quantal rotation has been recognized and investigated with a distinguished history for one hundred years. In nuclear physics, the rotation features of the nucleus were firstly noted by
Teller and Wheeler in 1938~\cite{Teller1938Phys.Rev.53.778}.
A more complete explanation of the nuclear rotation was due to Bohr and Mottelson by pointing out that
the rotation was a consequence of deformation~\cite{Bohr1951Phys.Rev.81.134} and by developing a model that combined the individual and collective motions of the nucleons~\cite{Bohr1975B}.

Since then it was generally accepted that the high angular momenta of atomic nuclei are connected with the collective rotations with a stable nuclear deformation. The nuclear rotational bands are built on the states with substantial quadrupole deformations and characterized by strong electric quadrupole (E2) transitions, which can be interpreted as the coherent collective rotation of many nucleons around an axis perpendicular to the symmetry axis of the nuclear density distribution~\cite{Bohr1975B}. The study along this line has been at the forefront of nuclear structure physics for several decades, as evidenced by the exciting discoveries such as the backbending~\cite{Johnson1971Phys.Lett.605},
angular momentum alignment~\cite{Stephens1972Nucl.Phys.257,Banerjee1973Nucl.Phys.366},
superdeformed rotational bands~\cite{Twin1986Phys.Rev.Lett.811}, etc.

In earlier 1990s, however, the rotational-like sequences of strongly enhanced magnetic dipole (M1)
transitions were surprisingly observed in several light-mass Pb isotopes, which are known to be
spherical or near-spherical. Sequentially,  this new type of rotational bands which have strong M1
and very weak E2 transitions has been well discovered experimentally in a number of nearly spherical nuclei (for reviews see Refs.~\cite{Hubel2005Prog.Part.Nucl.Phys.1,Frauendorf2001Rev.Mod.Phys.463,Clark2000Annu.Rev.Nucl.Part.Sci.1}).
Besides the outstanding magnetic dipole transitions, the intriguing feature here is that the orientation of the
rotor is not specified by the deformation of the overall density but rather by the current distribution
induced by specific nucleons moving in high-$j$ orbitals.

\begin{figure}[!htbp]
\centering
\includegraphics[width=7.0cm]{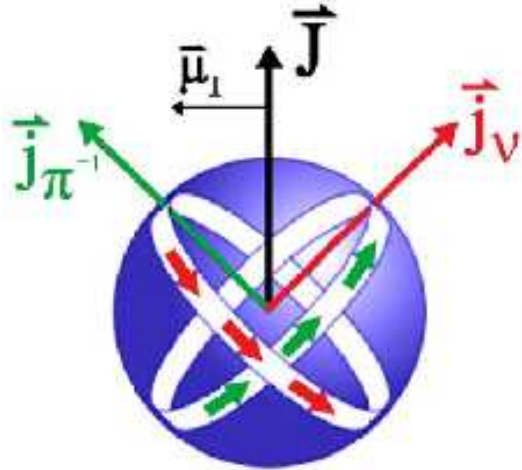}
\caption{A schematic illustration for the spin-coupling scheme of magnetic rotation. For a near-spherical nuclei,
the coupling of the proton-hole $j_{\pi}^{-1}$ and neutron-particle $j_{\nu}$, each in high-$j$ orbital,
gives the total angular momentum $J$. As a result, a large transverse component of the magnetic dipole moment vector, $\mu_\perp$, rotates around the total angular momentum vector, and creates the enhanced M1 transitions.}
\label{fig1-1}
\end{figure}

The explanation of such bands was given in terms of the shears mechanism~\cite{Frauendorf1993Nucl.Phys.A259}. In this interpretation, the magnetic
dipole vector, which arises from proton particles (holes) and
neutron holes (particles) in high-$j$ orbitals, rotates around the
total angular momentum vector. See Fig.~\ref{fig1-1} for a schematic illustration. At the bandhead, the proton and
neutron angular momenta are almost perpendicular to each other.
Along the bands, energy and angular momentum are increased by an
alignment of the proton and neutron angular momenta
along the total angular momentum. Consequently, the orientation of the
total angular momentum in the intrinsic frame does not change so
much and regular rotational bands are formed in spite of the fact
that the density distribution of the nucleus is almost spherical or
only weakly deformed. In order to distinguish this kind of rotation
from the usual collective rotation in well-deformed nuclei (called
electric rotation), the name ``magnetic rotation'' (MR) was introduced in
Ref.~\cite{Frauendorf199452}, which alludes to the fact that the magnetic moment is the
order parameter inducing a violation of rotational symmetry and thus causing rotational-like
structures in the spectrum~\cite{Frauendorf1997Z.Phys.A}. This forms an analogy to a ferromagnet
where the total magnetic moment, the sum of the atomic dipole moments, is the order parameter.

The experimental indicators for magnetic rotation can
be summarized as~\cite{Frauendorf2001Rev.Mod.Phys.463,Clark2000Annu.Rev.Nucl.Part.Sci.1}:

\noindent 1) a $\Delta I=1$ sequence of strong magnetic dipole transitions,
corresponding to a reduced transition probability $B(M1)\sim$ a few $\mu_N^2$;

\noindent 2) weak or absent quadrupole transitions, corresponding
to a deformation parameter $|\beta|\lesssim0.15$, which combined with strong M1 transitions
results in large $B(M1)/B(E2)$ ratios, $\gtrsim 20\mu_N^2/(eb)^{2}$;

\noindent 3) a smooth variation in the $\gamma$ transition energy with
angular momentum;

\noindent 4) a substantial dynamic moment of inertia, corresponding to the large ratio of the
${\cal J}^{(2)}/B(E2)\gtrsim 100 $MeV$^{-1}(eb)^{-2}$, compared with the values in well-deformed
[$\sim$10 MeV$^{-1}(eb)^{-2}$] or superdeformed [$\sim$5 MeV$^{-1}(eb)^{-2}$] rotational bands.

Similar as ferromagnetism and antiferromagnetism in condensed matter physics,
the existence of the antimagnetic rotation (AMR) in nuclear physics is also an interesting issue.
For an antiferromagnet, one-half of the atomic dipole moments are aligned on one sublattice and
the other half are aligned in the opposite direction on the other sublattice. In such way, the net magnetic
moment in an antiferromagnet is canceled out. However, it is still an ordered state since the isotropy
of such a state is also broken like a ferromagnet.

\begin{figure}[!htbp]
\centering
\includegraphics[width=8.0cm]{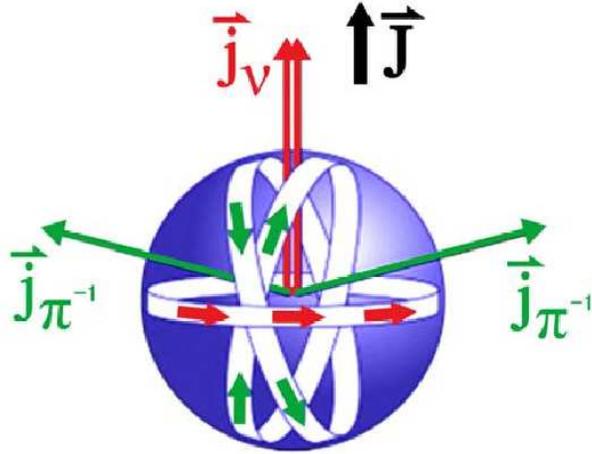}
\caption{A schematic illustration for the spin-coupling scheme of antimagnetic rotation.
Instead of the shears mechanism for magnetic rotation, the two shearslike configurations here result that
the magnetic moments are antialigned and cancel, and therefore the $B(M1)$ values vanish.}
\label{fig1-2}
\end{figure}

In analogy with an antiferromagnet, antimagnetic rotation~\cite{Frauendorf1996272,Frauendorf2001Rev.Mod.Phys.463} is predicted to occur in some specific nearly spherical nuclei, in which the subsystems of valence protons (neutrons) are aligned back to back in opposite directions and nearly perpendicular to the orientation of the total spin of the valence neutrons (protons).
Such arrangement of the proton and neutron angular momenta also breaks the rotational symmetry
in these nearly spherical nuclei and causes excitations with rotational character on top of this bandhead.
Along this band, energy and angular momentum are increased by simultaneous closing of the two
proton (neutron) blades toward the neutron (proton) angular momentum vector. Consequently,
a new kind of rotational bands in nearly spherical nuclei is formed showing some analogy with an antiferromagnet.
A schematic illustration for AMR is shown in Fig.~\ref{fig1-2}.

The experimental indicators for antimagnetic rotation can
be summarized as:

\noindent 1) a $\Delta I=2$ sequence with E2 transitions only, as the cancellation of the
magnetic moments leads to the absence of the M1 transition;

\noindent 2) weak E2 transitions corresponding to the small deformation parameter
$|\beta|<0.15$, and the $B(E2)$ values decreasing with spin;

\noindent 3) a smooth variation in the $\gamma$ transition energy with
angular momentum;

\noindent 4) a substantial dynamic moment of inertia, similar to the MR band, corresponding to the large ratio of
the ${\cal J}^{(2)}/B(E2) \gtrsim 100 $MeV$^{-1}(eb)^{-2}$.

The MR and AMR~\cite{Hubel2005Prog.Part.Nucl.Phys.1,Frauendorf2001Rev.Mod.Phys.463,Clark2000Annu.Rev.Nucl.Part.Sci.1}
discussed above are based on the assumption that the nucleus involved is not triaxially deformed. If the nucleus is
 triaxially deformed and the aplanar rotation is then allowed, another exotic phenomenon, nuclear chirality, may occur~\cite{Frauendorf1997Nucl.Phys.A131,Meng2010JPhysG.37.64025}.

\subsection{Experimental progress}

Following the observation of many long cascades of magnetic dipole
$\gamma$-ray transitions in the neutron deficient Pb nuclei in the early
1990s~\cite{Hubel1991,Baldsiefen1991,Fant1991J.Phys.G17.319,Hubel1992Prof.Part.Nucl.Phys.28.427,
Clark1992Phys.Lett.B247,Baldsiefen1992Phys.Lett.B252,Kuhnert1992Phys.Rev.C133}, several attempts~\cite{Wang1992PhysRevLett.69.1737,Hughes1993PhysRevC.48.R2135,Clark1994PhysRevC.50.84,
Neffgen1995NuclPhysA.595.499,Moore1995PhysRevC.51.115}
have been made to measure the lifetimes for the states in MR bands.
The high-accuracy lifetime measurements for four M1 bands in $^{198,199}\rm Pb$
performed with GAMMASPHERE provided a clear evidence for shears mechanism~\cite{Clark1997Phys.Rev.Lett.1868}. Additional evidence for the shears mechanism was provided by measuring the $g$-factor of a dipole
band in $^{193}$Pb~\cite{Chmel1997PhysRevLett.79.2002}. It is demonstrated that at the bandhead
the longitudinal component of the magnetic moment
$\mu_\parallel$ has the value expected for an opening angle of
90$^\circ$ of the two blades composed of the suggested particles
and holes.

\begin{figure*}[!htbp]
\centering
\includegraphics[width=12cm]{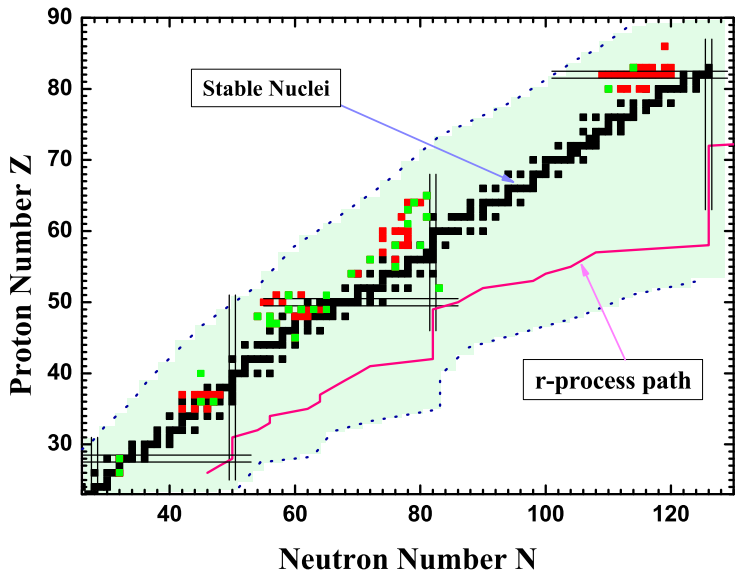}
\caption{The nuclides with magnetic rotation observed in the nuclear chart.
The red squares represented 56 nuclides with 120 magnetic dipole bands observed before 2000 and compiled in Ref.~\cite{Amita2000At.DataNucl.DataTables283}. The green squares represented the corresponding data after 2000.  }
\label{fig1-3}
\end{figure*}

From then on, more and more MR bands have been observed not only in the mass region of $A\sim190$ but also in
$A\sim80$, $A\sim110$, and $A\sim140$ regions. A compilation in these four mass regions up to December 2006 including 178 magnetic dipole bands in 76 nuclides is given in Ref.~\cite{Amita2000At.DataNucl.DataTables283} and additional data can be found in Refs.~\cite{Agarwal2007PhysRevC.76.024321,Yuan2007HyperfineInteractions.180.49,Bhattacharjee2009NuclPhysA.825.16,
Schwengner2009Phys.Rev.C44305,Deo2009Phys.Rev.C67304,Yuan2010ChinPhysB.19.062701,He2011Phys.Rev.C24309,Trivedi2012Phys.Rev.C14327,
Li2012NuclPhysA.892.34,Zhang2011PhysRevC.84.057302,Wang2012PhysRevC.86.064302,Ma2012EuroPhysJA.48.1}. Recent observations in $^{58}\rm Fe$~\cite{Steppenbeck2012Phys.Rev.C85}
and $^{60}\rm Ni$~\cite{Torres2008Phys.Rev.C54318} have extended the observed MR mass region to $A=60$. In total,  more than 195 magnetic dipole bands spread over 85 nuclides have been observed, which have been summarized in the nuclear chart in Fig.~\ref{fig1-3}.

It is difficult to define how much contribution from the collective rotation can be expected in the MR bands.
Obviously either the alignment of the angular momenta for the valent nucleons or the collective rotation due to deformation will cost energy. For a given angular momentum, the nuclear system will try to minimize its energy via a competition between the kinetic energy due to the collective rotation and the potential energy due to the closing of the blades of the shears blades. Weaker quadrupole deformation will lead to purer MR band and larger quadrupole deformation will reduce the $B(M1)/B(E2)$ ratios. Therefore it might be possible to observe the competition and transition between the  electric and the magnetic rotations. The experimental information for the lifetime measurement in $^{106}$Sn has suggested an extremely large ratio ${\cal J}^{(2)}/B(E2)>1000 $MeV$^{-1}(eb)^{-2}$, which may provide an example of almost pure magnetic rotation in a spherical nucleus~\cite{Jenkins1999PhysRevLett.83.500}.

AMR is expected to be observed in the same regions as MR in the nuclear chart~\cite{Frauendorf2001Rev.Mod.Phys.463}. However, it differs from magnetic rotation in two aspects. Firstly, there are no M1 transitions in the AMR band since the transverse magnetic moments of the two subsystems are antialigned and canceled out. Secondly, as the antimagnetic rotor is symmetric with respect to a rotation by $\pi$ about the rotating axis, the energy levels in AMR band differ in spin by $2\hbar$ and are connected by weak E2 transitions reflecting the nearly spherical core. Moreover, the AMR phenomenon is characterized by a decrease of the $B(E2)$ values with spin, which has been confirmed by lifetime measurements~\cite{Simons2003Phys.Rev.Lett.162501}. To date, AMR has attracted lots of attentions and has been observed in Cd isotopes including
$^{105}\rm Cd$~\cite{Choudhury2010Phys.Rev.C61308}, $^{106}\rm Cd$~\cite{Simons2003Phys.Rev.Lett.162501}, $^{108}\rm Cd$~\cite{Simons2005Phys.Rev.C24318,Datta2005Phys.Rev.C41305}, $^{109}\rm Cd$~\cite{Chiara2000Phys.Rev.C34318},
$^{110}\rm Cd$~\cite{Roy2011Phys.Lett.B322}, and $^{112}\rm Cd$~\cite{LiXW2012PhysRevC.86.057305}. The other candidates include $^{100}\rm Pd$~\cite{Zhu2001Phys.Rev.C41302}, and $^{144}\rm Dy$~\cite{Sugawara2009Phys.Rev.C64321}.

\subsection{Theoretical progress}

From the theoretical side, both the MR and AMR bands have been extensively discussed in recent years.
Using a semiclassical particle plus rotor model based on angular momentum geometry~\cite{Macchiavelli1998Phys.Rev.C3746,Macchiavelli1998Phys.Rev.C621,Clark2000Annu.Rev.Nucl.Part.Sci.1}, the competition between shears mechanism and core rotation has been investigated. As the mean field approach is easy
to construct classical vector diagrams showing
the angular momentum composition, it is widely used to understand the structure of these rotational bands.
Since for the magnetic rotation bands, the axis of the uniform rotation does not coincide with any principal axis of the
density distribution, a description of these bands requires a model going beyond the
principal axis cranking, which leads to the development of the tilted axis cranking (TAC) approach.

The semi-classical mean field approximation for tilted axis rotation can be traced back to the 1980s~\cite{Kerman1981Nucl.Phys.179,Frisk1987Phys.Lett.14}. After the first self-consistent TAC solutions were found in Ref.~\cite{Frauendorf1993Nucl.Phys.A259}, the qualities and interpretations of the TAC approximation were discussed and examined in Ref.~\cite{Frauendorf1996Z.Phys.A263} with the particle rotor model. Based on the TAC approximation, lots of applications are carried out in the framework of the pairing plus quadrupole model or the shell correction method~\cite{Frauendorf2000Nucl.Phys.A115,Frauendorf2001Rev.Mod.Phys.463}.

In the above mentioned TAC investigations, however, the polarization effects which are expected to strongly influence the quadrupole moments and thus the $B(E2)$ values, are either completely neglected or taken into account only partially by minimizing the rotating energy surface with respect to a few deformation parameters. Moreover, the nuclear currents, which are the origin of symmetry violation in nuclei with MR and/or AMR, are not treated in a self-consistent way in these models. Therefore, it is necessary to investigate the MR and AMR based on thoeries which describe the polarization effects and nuclear currents self-consistently without additional parameters. Such calculations are not simple, but they are nowadays feasible in the framework of density functional theory (DFT).

The DFT with a small number of parameters allows a very
successful description of ground-state and excited-state properties of nuclei all over the nuclear
chart. On the basis of the density functionals, the rotational excitations have been described in the
principal-axis cranking Hartree-Fock (HF) or Hartree-Fock-Bogoliubov (HFB) framework with the zero-range Skyrme force~\cite{Fleckner1979NuclPhysA.331.288, Flocard1982NuclPhysA.391.285, Bonche1987NuclPhysA.467.115} as well as the density dependent Gogny force~\cite{Egido1993PhysRevLett.70.2876} non-relativistically.
The three-dimensional TAC version, which allows to study nuclear aplanar rotation, has been developed with the Skyrme density functional and
applied to not only the magnetic rotation in $^{142}$Gd in Ref.~\cite{Olbratowski2002APPB.33.389}
but also chirality~\cite{Frauendorf1997Nucl.Phys.A131} in $A\sim130$ mass regions~\cite{Olbratowski2004PhysRevLett.93.052501, Olbratowski2006PhysRevC.73.054308}.

The covariant version of DFT takes the Lorentz symmetry into account
in a self-consistent way and has received wide attention due to
its successful description for lots of nuclear phenomena in stable as well as exotic
nuclei~\cite{Ring1996Prog.Part.Nucl.Phys.193,Vretenar2005Phys.Rep.101,Meng2006Prog.Part.Nucl.Phys.470}. The representations with
large scalar and vector fields in nuclei, of the order of a few hundred MeV, provide more efficient
descriptions than non-relativistic approaches that hide these scales. It can obtain reasonable nuclear saturation properties in infinite nuclear
matter with Brueckner method~\cite{Brockmann1990PhysRevC.42.1965,Brockmann1992PhysRevLett.68.3408}, reproduce well the measurements of the isotopic
shifts in the Pb region~\cite{Sharma1993PhysLettB.317.9}, give naturally the spin-orbit potential, explain the origin of the pseudospin symmetry~\cite{Arima1969PhysLettB.30.517,Hecht1969NuclPhysA.137.129} as a relativistic symmetry~\cite{Ginocchio1997PhysRevLett.78.436,Meng1998PhysRevC.58.R628,Meng1999PhysRevC.59.154,
Long2006PhysLettB.639.242,Liang2011PhysRevC.83.041301,Guo2012PhysRevC.85.021302,Lu2012PhysRevLett.109.072501} and the spin
symmetry in the anti-nucleon spectrum~\cite{Zhou2003PhysRevLett.91.262501,Liang2010EuroPhysJA.44.119}, and is reliable for nuclei far away from the $\beta$-stability line~\cite{Zhao2010Phys.Rev.C54319,Zhao2012Phys.Rev.C64324}, etc.
Moreover, it is of particular importance that the covariant density functional theory (CDFT) includes nuclear magnetism~\cite{Koepf1989Nucl.Phys.A61}, and provides a consistent description of
currents and time-odd fields, which plays an important role in the nuclear rotations.

Based on the CDFT, Koepf and Ring~\cite{Koepf1989Nucl.Phys.A61,Koepf1990Nucl.Phys.A279} first developed the principal axis cranking relativistic Hartree calculations. K\"{o}nig and Ring~\cite{Konig1993PhysRevLett.71.3079} included the effects of isoscalar and isovector baryon currents that are the sources of the magnetic potentials in the Dirac equation. The nuclear magnetism appeared to be crucial for the quantitative understanding of nuclear
moment of inertia. Large scale cranking relativistic Hartree-Bogoliubov calculations using the central part of the Gogny force with the D1S parameter set acting in the pairing channel has been performed and summarized in Ref.~\cite{Afanasjev1999PhysRevC.60.051303}. These calculations went beyond the mean-field by
invoking an approximate number projection which
appears to have great impact on the ${\cal J}^{(2)}$ values.

The three-dimensional cranking CDFT has been developed in
Ref.~\cite{Madokoro2000Phys.Rev.C61301}, in which
the nucleon and meson fields are expanded in terms of three-dimensional
harmonic-oscillator eigenfunctions.
In comparison with the principal axis cranking (PAC) case, the signature symmetry is broken
and the parity is thus the only symmetry left~\cite{Madokoro2000Phys.Rev.C61301}.
Even without the pairing correlations, it is
very time consuming to perform a three-dimensional cranking calculation.
Therefore the three-dimensional cranking CDFT is firstly used to examine its applicability to the shears bands, i.e.,
the magnetic rotation in $^{84}\rm Rb$~\cite{Madokoro2000Phys.Rev.C61301}.

Focusing on the magnetic rotational bands, a completely new computer code
for the self-consistent two-dimensional cranking CDFT based on the non-linear meson-exchange interaction  has been
established~\cite{Peng2008Phys.Rev.C24313}. This new code includes significant improvements, such as implanting the
simplex symmetry ${\cal P}_y{\cal T}$, quantum number transformation,
and orientation constraints. Compared with the three-dimensional cranking CDFT,
the computing time for this two-dimensional cranking CDFT model has been considerably reduced and thus it allows systematic investigations and has been applied to heavy nucleus like $^{142}$Gd~\cite{Peng2008Phys.Rev.C24313}.

The CDFT with the point-coupling interaction become very popular in recent years owing to the following advantages, 1) it avoids the possible physical constrains introduced by explicit usage of the Klein-Gordon equation, especially the fictitious $\sigma$ meson; 2) it is possible to study the naturalness of the interaction~\cite{Friar1996PhysRevC.53.3085,Manohar1984NuclPhysB.234.189} in effective theories for related nuclear structure problems; 3) it is relatively easy to include the Fock terms~\cite{Liang2012PhysRevC.86.021302} and to investigate its relationship to the nonrelativistic approaches~\cite{Sulaksono2003AnnPhys.308.354}.

The TAC model based on the CDFT with the point-coupling interaction was established in Ref.~\cite{Zhao2011Phys.Lett.B181}, which introduces further simplification and reduces computing time.
It has been applied successfully for the MR ranging from light nuclei such as $^{60}\rm Ni$~\cite{Zhao2011Phys.Lett.B181} and $^{58}$Fe~\cite{Steppenbeck2012Phys.Rev.C85}, to medium heavy nuclei such as $^{114}$In~\cite{Li2012NuclPhysA.892.34}, and to heavy nuclei such as $^{198,199}\rm Pb$~\cite{Yu2012Phys.Rev.C24318}.
It also provides a fully self-consistent and microscopic investigation for the
AMR in $^{105}\rm Cd$~\cite{Zhao2011Phys.Rev.Lett.122501}. More examples of its application for the AMR bands in $^{112}$In can be found in Ref.~\cite{LiXW2012PhysRevC.86.057305}.

In the review, we will mainly focus on the microscopic CDFT investigation for the MR and AMR. In Section II, the  theoretical framework of the tilted axis cranking CDFT  with the point-coupling interaction will be
introduced. In Section III, taking the seminal MR in $^{198}$Pb as an example,
the CDFT description for the MR including the single particle Routhian, energy spectra, deformation evolution, shears mechanism, electric and magnetic transitions properties, etc. will be outlined.
Similar applications for MR in the mass regions $A\sim60, 80, 100, 140$ are summarized in the next Section.
In Section V, taking $^{105}$Cd as an example, the CDFT investigation for the AMR are discussed. Finally, the summary and perspectives are given in Section VI.

\section{Tilted axis cranking covariant density functional theory}

The basic idea of the cranking model is based on the following classical assumption~\cite{Ring1980}:
 If one introduces a coordinate system which rotates with constant angular velocity around a fixed axis in space, the motion of the nucleons in the rotating frame is rather simple if the angular frequency is properly chosen. In particular, the nucleons can be thought of as independent particles moving in an average potential which is rotating with the coordinate frame.

The advantages of the cranking model include, 1) providing a microscopic description for rotating nucleus;
2) describing the collective angular momentum as the sum of the single particle ones; 3) working extremely well at high angular momentum where the assumption of uniform rotation applies.

For the magnetic rotations, the axis of the uniform rotation does
not coincide with any principal axis of the density distribution, a description of the  $\Delta I=1$ rotational bands
requires going beyond the principal axis cranking, which leads to the TAC approach~\cite{Frauendorf1993Nucl.Phys.A259}.
The initial version of the TAC approach was developed based on the  potentials of the Nilsson type, which are combined with a pairing plus quadrupole model or the shell correction method for finding the deformation.

In this section, we first present the basic idea of TAC approach by taking the simple pairing
plus quadrupole model (PQTAC) as an example. The details of the PQTAC model can be found in Refs.~\cite{Frauendorf1993Nucl.Phys.A259,Frauendorf2000Nucl.Phys.A115,Frauendorf2001Rev.Mod.Phys.463}.
In Subsection~\ref{Subsec2.2}, the theoretical framework of the covariant density functional theory
will be briefly introduced. The combination of the TAC approach and the covariant density functional theory, i.e., the so-called tilted axis cranking relativistic mean-field theory, is given in the Subsection~\ref{Subsec2.3}. Finally, Subsection~\ref{Subsec2.4} will be devoted to the numerical techniques.

 \subsection{Tilted axis cranking model}\label{Subsec2.1}

Assuming that a nucleus rotates around an arbitrary axis with a uniform velocity $\bm{\Omega}$, the corresponding Hamiltonian (with the pairing plus quadrupole interaction) in intrinsic frame, i.e., the Routhian, can be written as,
 \begin{equation}\label{Eq.1}
  H'= H - \bm{\Omega}\cdot\bm{J} = H_{sph}-\frac{\chi}{2}\sum\limits_{\mu=-2}^2Q^\dagger_\mu Q_\mu-GP^\dagger P-\lambda \hat{N}- \bm{\Omega}\cdot\bm{J}.
 \end{equation}
The constraint $\bm{\Omega}\cdot\bm{J}$ ensures that the rotational states have a finite angular momentum $\bm{J}$. The justification of pairing plus quadrupole Hamiltonian is described in many textbooks (e.g., Ref.~\cite{Ring1980}). In PQTAC model, $H_{sph}$ is approximated by the spherical part of the standard Nilsson Hamiltonian. The residual interaction includes two parts. The quadrupole interaction defined by the quadrupole operators is responsible for the quadrupole deformation of the mean field. The pairing interaction takes a monopole form. The term $\lambda \hat{N}$ controls the particle number $N$. The Hamiltonian (\ref{Eq.1}) is written only for one kind of nucleons. The terms $H_{sph}$ and $Q_\mu$ should be understood as sums of a proton and a neutron part, and there are terms $-GP^\dagger P$ and $-\lambda \hat{N}$ for both protons and neutrons. In practice, the actual values of the force constants $\chi$ and $G$ depend on the configuration space under consideration and are determined from experimental data.

Approximating the nuclear wave function as a Slater determinant in the quasiparticle space and neglecting the exchange terms, the single particle Routhian $h'$ becomes
 \begin{equation}\label{Eq.2}
    h' = h_{sph}-\frac{1}{2}\sum\limits_{\mu=-2}^2 \left( q_\mu Q^\dagger_\mu +q_\mu^\ast Q_\mu\right)-\Delta
                  \left(P^\dagger+P\right)-\lambda \hat{N}- \bm{\Omega}\cdot\bm{J}
 \end{equation}
with the deformed potential $q_\mu=\chi\langle Q_\mu\rangle$ and the pairing potential $\Delta=G\langle P\rangle$. The Fermi surface $\lambda$ is determined by $N=\langle\hat{N}\rangle$, with $N$ the particle number of neutrons or protons.

For given parameters $q_\mu$ and $\Delta$, solving Eq.~(\ref{Eq.2}) is similar to solving a cranking Nilsson Hamiltonian. Here $q_\mu$ and $\Delta$ depend on the wave function and are determined by iteration, or equivalently by minimizing the total Routhian $E'(q_\mu,\Delta)=\langle H'\rangle$ for each rotational frequency $\Omega$. It has been shown in Ref.~\cite{Kerman1981Nucl.Phys.179} that a self-consistent solution is equivalent to the requirement that the vector of the rotational frequency $\bm{\Omega}$ is parallel to the vector of the angular momentum. Finally, the total energy as function of the angular momentum is obtained by
 \begin{equation}
 E(I) = E'(\Omega)+\Omega I(\Omega),
 \end{equation}
where $I(\Omega)$ is determined with $I=|\langle\bm{J}\rangle|$.

 Usually, it is convenient to choose the principal axes of the density distribution as the axes of the coordinate frame, i.e.,
 \begin{equation}
  q_{-1}=q_{1}=0,\quad\quad q_{-2}=q_2.
 \end{equation}
In this case, the deformation of the potential is specified by two intrinsic quadrupole moments $q_0$ and $q_2$. Therefore, it is useful to reformulate the TAC model in this intrinsic frame~\cite{Frauendorf1993Nucl.Phys.A259,Frauendorf2000Nucl.Phys.A115,Frauendorf2001Rev.Mod.Phys.463}.

Cranking model are based on the classical treatment of the total angular momentum and the assumption of uniform rotation, which have the consequence that angular momentum conservation is violated. The connection with the quantal spectra is made by means of semiclassical expressions for the energy and transition matrix elements. Hence, it is important to investigate how well these approximations work for the description of the experimental observables. Moreover, in the TAC approach, the rotational axis does not coincides with any principal axis, i.e., it is tilted away from the principal axis. Different from the principal axis cranking model, the signature is no longer a good quantum number in the TAC. This leads to a problem of how to interpret the TAC solutions and construct the excitation spectrum from the TAC quasiparticle levels avoiding spurious states.

 The above questions have been discussed in Ref.~\cite{Frauendorf1996Z.Phys.A263} by comparing with the particle rotor model, which treats the quantal angular momentum dynamics properly. It was found that the TAC approach quantitatively accounts both for the energies and the intra band transition rates of the lowest bands generated by one or two quasi particles coupled to an axial rotor. The TAC provides an accurate description of the bandhead, except in cases, when substantial alignment of quasi particle angular momentum occurs at very low frequency.

The TAC approach is based on the mean-field theory. Therefore, one may easily study multi-quasiparticle excitations, and the consequences of changes of the deformation or the pairing. It gives transparent classical vector diagrams pictures of the angular momentum coupling, which is of great help to understand the structure of rotation bands.
The drawback of the TAC model is that it cannot describe the gradual onset of signature splitting
as well as the mixing of bands with substantially different quasiparticle angular momentum as in the standard cranking theory.

 \subsection{Covariant density functional theory}\label{Subsec2.2}

The covariant density functional theory can be traced back to the successful relativistic mean-field (RMF) models introduced by Walecka and Serot~\cite{Serot1986Adv.Nucl.Phys.1,Serot1997Int.J.Mod.Phys.E515}, which was further developed and applied by many groups~\cite{Ring1996Prog.Part.Nucl.Phys.193,Vretenar2005Phys.Rep.101,Meng2006Prog.Part.Nucl.Phys.470}. The most popular RMF models are based on the finite-range meson-exchange representation, in which the nucleus is described as a system of Dirac nucleons that interact with each other via the exchange of mesons. The nucleons and the mesons are described by the Dirac equation and the Klein-Gordon equation, respectively. The isoscalar-scalar $\sigma$ meson, the isoscalar-vector $\omega$ meson, and the isovector-vector $\rho$ meson build the minimal set of meson fields that, together with the electromagnetic field, is necessary for a description of bulk and single-particle nuclear properties. Moreover, a quantitative treatment of nuclear matter and finite nuclei needs a medium dependence of effective mean-field interactions, which can be introduced by including nonlinear meson self-interaction terms in the Lagrangian or by assuming explicit density dependence for the meson-nucleon couplings. The detailed formulism of the meson-exchange representation of CDFT can be found in Refs.~\cite{Ring1996Prog.Part.Nucl.Phys.193,Vretenar2005Phys.Rep.101,Meng2006Prog.Part.Nucl.Phys.470}.

  More recently, this framework has been reinterpreted by the relativistic Kohn¨CSham density functional theory, and the functionals have been developed based on the zero-range point-coupling interaction~\cite{Nikolaus1992Phys.Rev.C1757,Burvenich2002Phys.Rev.C44308,Zhao2010Phys.Rev.C54319}, in which the meson exchange in each channel (scalar-isoscalar, vector-isoscalar, scalar-isovector, and vector-isovector) is replaced by the corresponding local four-point (contact) interaction between nucleons. In recent years, the point-coupling model has attracted more and more attention owing to the following advantages. First, it avoids the possible physical constrains introduced by explicit usage of the Klein-Gordon equation to describe mean meson fields, especially the fictitious $\sigma$ meson. Second, it is possible to study the role of naturalness~\cite{Friar1996PhysRevC.53.3085,Manohar1984NuclPhysB.234.189} in effective theories for nuclear-structure-related problems. Third, it is relatively easy to include the Fock terms~\cite{Liang2012PhysRevC.86.021302}, and provides more opportunities to investigate its relationship to the nonrelativistic approaches~\cite{Sulaksono2003AnnPhys.308.354}. In the following, we present the theoretical framework of the CDFT with the point-coupling interaction.

  The basic building blocks of the covariant density functional theory with point-coupling vertices are
  \begin{equation}
    (\bar\psi{\cal O}\Gamma\psi),~~~~~{\cal O}\in\{1,\vec{\tau}\},~~~~~\Gamma\in\{1,\gamma_\mu,\gamma_5,    \gamma_5\gamma_\mu,\sigma_{\mu\nu}\},
  \end{equation}
  where $\psi$ is Dirac spinor field of nucleon, $\vec{\tau}$ is the isospin Pauli matrix, and $\Gamma$ generally denotes the $4\times4$ Dirac matrices. There are ten such building blocks characterized by their transformation characteristics in isospin and Minkowski space. In the following, the vectors in the isospin space are denoted by arrows and the space vectors by bold type. Greek indices $\mu$ and $\nu$ run over the Minkowski indices $0$, $1$, $2$, and $3$.

  A general effective Lagrangian can be written as a power series in $\bar\psi{\cal O}\Gamma\psi$ and their derivatives, with higher-order terms representing in-medium many-body correlations. In the actual application we start with the following Lagrangian density of the form:
  \begin{eqnarray}\label{EQ:LAG}
  {\cal L}&=&{\cal L}^{\rm free}+{\cal L}^{\rm 4f}+{\cal L}^{\rm hot}+{\cal L}^{\rm der}+{\cal L}^{\rm em} \nonumber\\
          &=&\bar\psi(i\gamma_\mu\partial^\mu-m)\psi \nonumber\\
          &&-\frac{1}{2}\alpha_S(\bar\psi\psi)(\bar\psi\psi)
            -\frac{1}{2}\alpha_V(\bar\psi\gamma_\mu\psi)(\bar\psi\gamma^\mu\psi)
            -\frac{1}{2}\alpha_{TS}(\bar\psi\vec{\tau}\psi)(\bar\psi\vec{\tau}\psi)
            -\frac{1}{2}\alpha_{TV}(\bar\psi\vec{\tau}\gamma_\mu\psi)(\bar\psi\vec{\tau}\gamma^\mu\psi) \nonumber\\
         &&-\frac{1}{3}\beta_S(\bar\psi\psi)^3-\frac{1}{4}\gamma_S(\bar\psi\psi)^4-\frac{1}{4}
           \gamma_V[(\bar\psi\gamma_\mu\psi)(\bar\psi\gamma^\mu\psi)]^2
           -\frac{1}{2}\delta_S\partial_\nu(\bar\psi\psi)\partial^\nu(\bar\psi\psi)\nonumber\\
         &&-\frac{1}{2}\delta_V\partial_\nu(\bar\psi\gamma_\mu\psi)\partial^\nu(\bar\psi\gamma^\mu\psi)
           -\frac{1}{2}\delta_{TS}\partial_\nu(\bar\psi\vec\tau\psi)\partial^\nu(\bar\psi\vec\tau\psi)
           -\frac{1}{2}\delta_{TV}\partial_\nu(\bar\psi\vec\tau\gamma_\mu\psi)\partial^\nu(\bar\psi\vec\tau\gamma_\mu\psi)\nonumber\\
         &&-\frac{1}{4}F^{\mu\nu}F_{\mu\nu}-e\frac{1-\tau_3}{2}\bar\psi\gamma^\mu\psi A_\mu,
  \end{eqnarray}
  which includes the Lagrangian density for free nucleons ${\cal L}^{\rm free}$, the four-fermion point-coupling terms ${\cal L}^{\rm 4f}$, the higher order terms ${\cal L}^{\rm hot}$ accounting for the medium effects, the derivative terms ${\cal L}^{\rm der}$ to simulate the effects of finite-range which are crucial for a quantitative description for nuclear density distributions (e.g., nuclear radii), and the electromagnetic interaction terms ${\cal L}^{\rm em}$. The higher order terms lead in the mean field approximation to density dependent coupling constants with a density dependence of polynomial form.

  For the Lagrangian density in Eq.~(\ref{EQ:LAG}), $m$ is the nucleon mass and $e$ is the charge unit for protons. $A_\mu$ and $F_{\mu\nu}$ are respectively the four-vector potential and field strength tensor of the electromagnetic field. There are totally 11 coupling constants, $\alpha_S$, $\alpha_V$, $\alpha_{TS}$, $\alpha_{TV}$, $\beta_S$, $\gamma_S$, $\gamma_V$, $\delta_S$, $\delta_V$, $\delta_{TS}$, and $\delta_{TV}$, in which $\alpha$ refers to the four-fermion term, $\beta$ and $\gamma$ respectively the third- and fourth-order terms, and $\delta$ the derivative couplings. The subscripts $S$, $V$, and $T$ respectively indicate the symmetries of the couplings, i.e., $S$ stands for scalar, $V$ for vector, and $T$ for isovector. The isovector-scalar channel including the terms $\alpha_{TS}$ and $\delta_{TS}$ in Eq.~(\ref{EQ:LAG}) is usually neglected since a fit including the isovector-scalar interaction does not improve the description of nuclear ground-state properties~\cite{Burvenich2002Phys.Rev.C44308}. Furthermore, the pseudoscalar $\gamma_5$ and pseudovector $\gamma_5\gamma_\mu$ channels are also neglected in Eq.~(\ref{EQ:LAG}) since they do not contribute at the Hartree level due to parity conservation in nuclei.

  The Hamiltonian density, i.e., the $00$ components of the energy-momentum tensor can be obtained by the Legendre transformation
  \begin{equation}
   {\cal H} = T^{00} = \frac{\partial{\cal L}}{\partial\dot\phi_i}\dot\phi_i - {\cal L},
  \end{equation}
  where $\phi_i$ represents the nucleon or photon field. Therefore, the total Hamiltonian reads
  \begin{eqnarray}
   H &=& \int d^3x{\cal H} \nonumber \\
     &=& \int d^3x\left\{\bar\psi\left[-i\bm{\gamma}\cdot\bm{\nabla} + m\right]\psi\right. \nonumber \\
     &&+\frac{1}{2}\alpha_S(\bar\psi\psi)(\bar\psi\psi)+\frac{1}{2}\alpha_V(\bar\psi\gamma_\mu\psi)(\bar\psi\gamma^\mu\psi)
       +\frac{1}{2}\alpha_{TV}(\bar\psi\vec{\tau}\gamma_\mu\psi)(\bar\psi\vec{\tau}\gamma^\mu\psi)\nonumber \\
     &&+\frac{1}{3}\beta_S(\bar\psi\psi)^3+\frac{1}{4}\gamma_S(\bar\psi\psi)^4+\frac{1}{4}  \gamma_V[(\bar\psi\gamma_\mu\psi)(\bar\psi\gamma^\mu\psi)]^2\nonumber \\
     &&-\frac{1}{2}\delta_S\left[\partial_0(\bar\psi\psi)\partial^0(\bar\psi\psi)+\bm{\nabla}(\bar\psi\psi)\cdot\bm{\nabla}(\bar\psi\psi)\right]\nonumber \\
     &&-\frac{1}{2}\delta_V\left[\partial_0(\bar\psi\gamma_\mu\psi)\partial^0(\bar\psi\gamma^\mu\psi)+\bm{\nabla}(\bar\psi\gamma_\mu\psi)\cdot\bm{\nabla}(\bar\psi\gamma^\mu\psi)\right]\nonumber  \\
     &&-\frac{1}{2}\delta_{TV}\left[\partial_0(\bar\psi\vec{\tau}\gamma_\mu\psi)\partial^0(\bar\psi\vec{\tau}\gamma^\mu\psi)+\bm{\nabla}(\bar\psi\vec{\tau}\gamma_\mu\psi)\cdot\bm{\nabla}(\bar\psi\vec{\tau}\gamma^\mu\psi)\right] \nonumber  \\
     &&\left.+e\frac{1-\tau_3}{2}\bar\psi\gamma^\mu A_\mu\psi-F^{0\mu}\partial^0A_\mu+\frac{1}{4}F_{\mu\nu}F^{\mu\nu}\right\}.
  \end{eqnarray}

  One can describe the nucleon field as
  \begin{subequations}
   \begin{eqnarray}
     \psi(x)&=&\sum_k \left[ \psi_k(x)  a_k + \psi^{'}_{k}(x)d_{k}^{\dag}\right],\\
     \psi^\dag(x)&=&\sum_k \left[ \psi^\dag_k(x) a_k^\dag + \psi^{'\dag}_{k}(x)d_{k}\right],
   \end{eqnarray}
  \end{subequations}
  where $\psi_k(x)$ and $\psi^{'}_{k}(x)$ construct a complete set of the Dirac spinor. The operators $a_{k}$ and $a_k^\dag$ respectively represent the creation and annihilate operators for nucleons, whereas $d_{k}$ and $d_k^\dag$ represent those for the antinucleons which corresponds to the negative energy states in the Dirac sea. Usually the contributions from the Dirac sea are neglected in the framework of CDFT, i.e., the so-called ``no-sea'' approximation~\cite{Serot1986Adv.Nucl.Phys.1}.

The nuclear ground-state wavefunction $\vert\Phi\rangle$ is assumed as a Slater determinant
  \begin{equation}\label{IntrinsicWF}
   \vert\Phi\rangle=\prod\limits_{k=1}^Aa^\dag_k\vert-\rangle,
  \end{equation}
  where $\vert-\rangle$ is the physical vacuum. This leads to the replacement of the operators $\bar\psi(\hat{{\cal O}}\Gamma)_i\psi$ in Eq. (\ref{EQ:LAG}) by their expectation values which become bilinear forms of the nucleon Dirac spinor $\psi_k$ in the Hartree approximation,
  \begin{equation}
   \bar\psi(\hat {\cal O}\Gamma)_i \psi\rightarrow\langle \Phi\vert :\bar\psi(\hat {\cal O}\Gamma)_i \psi:\vert\Phi\rangle
   =\sum_{k>0}\bar\psi_k(x)(\hat {\cal O}\Gamma)_i \psi_k(x),
  \end{equation}
where $i$ indicates $S$, $V$, and $TV$, and the sum $\sum$ runs over only positive energy states. Accordingly, the energy density functional for a nuclear system can be represented as,
  \begin{eqnarray} \label{Energy}
  E_{\rm CDF}&\equiv&\langle\Phi\vert H\vert\Phi\rangle \nonumber \\
            &=&\int d^3x\left\{\sum_{k=1}^A{\psi^\dagger_k \left(\bm{\alpha}\cdot\bm{p} + \beta m\right)\psi_k}\right.
            +\frac{1}{2}\alpha_S\rho_S^2+\frac{1}{2}\alpha_V j_\mu j^\mu+\frac{1}{2}\alpha_{TV}(\vec j_{TV})_\mu\vec j^\mu_{TV}\nonumber \\
            &&+\frac{1}{3}\beta_S\rho_S^3+\frac{1}{4}\gamma_S\rho_S^4+\frac{1}{4}\gamma_V(j_{\mu}j^\mu)^2-\frac{1}{2}\delta_S\left[\partial_0\rho_S\partial^0\rho_S+\bm{\nabla}\rho_S\cdot\bm{\nabla}\rho_S\right]\nonumber \\
            &&-\frac{1}{2}\delta_V\left[\partial_0j_{\mu}\partial^0j^\mu+\bm{\nabla}j_{\mu}\cdot\bm{\nabla}j^\mu\right]-\frac{1}{2}\delta_{TV}\left[\partial_0(\vec j_{TV})_\mu\partial^0\vec j^\mu_{TV}+\bm{\nabla}(\vec j_{TV})_\mu\cdot\bm{\nabla}\vec j^\mu_{TV}\right] \nonumber  \\
            &&\left.+eA_\mu j_p^\mu-F^{0\mu}\partial^0A_\mu+\frac{1}{4} F_{\mu\nu}F^{\mu\nu}\right\},
  \end{eqnarray}
  where the local densities and currents read,
  \begin{subequations}\label{currents}
   \begin{equation}
    \rho_S(x)=\langle \Phi\vert :\bar\psi \psi:\vert\Phi\rangle=\sum_{k=1}^A\bar\psi_k(x)\psi_k(x),
   \end{equation}
   \begin{equation}
    j^\mu(x)=\langle \Phi\vert:\bar\psi\gamma^\mu \psi:\vert\Phi\rangle=\sum_{k=1}^A\bar\psi_k(x)\gamma^\mu\psi_k(x),
   \end{equation}
   \begin{equation}
    \vec j^{\mu}_{TV}(x)=\langle \Phi\vert:\bar\psi\gamma^\mu\vec\tau \psi:\vert\Phi\rangle =\sum_{k=1}^A\bar\psi_k(x)\gamma^\mu\vec\tau\psi_k(x),
   \end{equation}
   \begin{equation}
    j^\mu_p(x)=\langle \Phi\vert:\bar\psi\gamma^\mu\frac{1-\tau_3}{2} \psi:\vert\Phi\rangle            =\sum_{k=1}^A\bar\psi_k(x)\gamma^\mu\frac{1-\tau_3}{2}\psi_k(x).
   \end{equation}
  \end{subequations}

  For the stationary case,
   $\psi_k(x) = \psi_k(\bm{r})e^{-i\varepsilon_kt}$,
  by minimizing the energy density functional Eq.~(\ref{Energy}) with respect to $\bar\psi_k$, one obtains the Dirac equation for the single nucleons
   \begin{equation}\label{Dirac}
     [\bm{\alpha}\cdot(-i\bm{\nabla}-\bm{V})+ V +\beta(m+S)]\psi_k=\varepsilon_k\psi_k.
   \end{equation}
  Here, the local scalar $S(\bm{r})$ and vector $V^\mu(\bm{r})$ potentials read
   \begin{equation}\label{potential}
     S(\bm{r})    =\Sigma_S, \quad  V^\mu(\bm{r})=\Sigma^\mu+\vec\tau\cdot\vec\Sigma^\mu_{TV},
   \end{equation}
  where the nucleon scalar-isoscalar $\Sigma_S$, vector-isoscalar $\Sigma^\mu$, and vector-isovector $\vec\Sigma^\mu_{TV}$ self-energies are given in terms of the various densities,
  \begin{subequations}
    \begin{eqnarray}
     \Sigma_S           &=&\alpha_S\rho_S+\beta_S\rho^2_S+\gamma_S\rho^3_S+\delta_S\triangle\rho_S,\\
     \Sigma^\mu         &=&\alpha_Vj^\mu +\gamma_V (j^\mu)^3 +\delta_V\triangle j^\mu + e \frac{1-\tau_3}{2}A^\mu,\\
     \vec\Sigma^\mu_{TV}&=& \alpha_{TV}\vec j^\mu_{TV}+\delta_{TV}\triangle\vec j^\mu_{TV}.
    \end{eqnarray}
  \end{subequations}
  Similarly, one can also obtain the Coulomb field $A^\mu$ which is determined by Poisson's equation
  \begin{equation}
    -\triangle A^\mu(\bm{r}) = ej^\mu_p(\bm{r}).
  \end{equation}

  For a system with time reversal invariance, the space-like components of the currents $\bm{j}_i$ in Eq.~(\ref{currents}) and the vector potential $\bm{V}(\bm{r})$ in Eq.~(\ref{potential}) vanish. Further assuming that the nucleon single-particle states do not mix isospin, i.e., the single-particle states are eigenstates of $\tau_3$, only the third component of isovector potentials $\vec\Sigma^\mu_{TV}$ survives. The total energy is thus given by
  \begin{eqnarray}
    E_{\rm CDF}&=&\int d^3x\left\{\sum_{k=1}^A{\psi^\dagger_k \left(\bm{\alpha}\cdot\bm{p} + \beta m\right)\psi_k}\right.
            +\frac{1}{2}\alpha_S\rho_S^2+\frac{1}{2}\alpha_V \rho_V^2+\frac{1}{2}\alpha_{TV}\rho_{TV}^2\nonumber \\
            &&+\frac{1}{3}\beta_S\rho_S^3+\frac{1}{4}\gamma_S\rho_S^4+\frac{1}{4}\gamma_V\rho_V^4
            +\frac{1}{2}\delta_S\rho_S\triangle\rho_S\nonumber \\
            &&+\frac{1}{2}\delta_V\rho_V\triangle\rho_V
            \left.+\frac{1}{2}\delta_{TV}\rho_{TV}\triangle\rho_{TV}+\frac{1}{2}eA_0 \rho_p\right\}.
  \end{eqnarray}

  As the translational symmetry is broken in the mean-field approximation, the center-of-mass (c.m.) correction should be made for the spurious c.m. motion. Nowadays, this is usually done by including the microscopic c.m. correction energy~\cite{Bender2000Euro.Phys.J.A467,Zhao2009Chin.Phys.Lett.112102}
  \begin{equation} \label{Eq:Ecm}
   E^{\rm mic}_{\rm c.m.}=-\frac{1}{2mA}\langle\hat{\bm P}^{2}_{\rm c.m.}\rangle,
  \end{equation}
  with $A$ being the mass number and $\hat{\bm{P}}_{\rm c.m.}=\sum_i^A \hat{\bm{p}}_i$ being the total momentum in the c.m. frame. Therefore, the total energy for the nuclear system becomes
  \begin{equation}
    E_{\rm tot} = E_{\rm CDF}+E^{\rm mic}_{\rm c.m.}.
  \end{equation}

  In the framework of the CDFT with the point-coupling interaction, a new point-coupling effective interaction PC-PK1 has been proposed by fitting to observables of 60 selected spherical nuclei, including the binding energies, charge radii, and empirical pairing gaps~\cite{Zhao2010Phys.Rev.C54319}. This effective interaction particularly improves the description for isospin dependence of binding energies and it has been successfully used in describing the Coulomb displacement energies between mirror nuclei~\cite{Sun2011Sci.ChinaSer.G-Phys.Mech.Astron.210}, fission barriers~\cite{Lu2012Phys.Rev.C11301}, the new and accurate mass measurement results at Gesellschaft f\"{u}r Schwerionenforschung (GSI)~\cite{Zhao2012Phys.Rev.C64324}, etc.

\begin{figure}[!htbp]
  \centering
  \includegraphics[width=9cm]{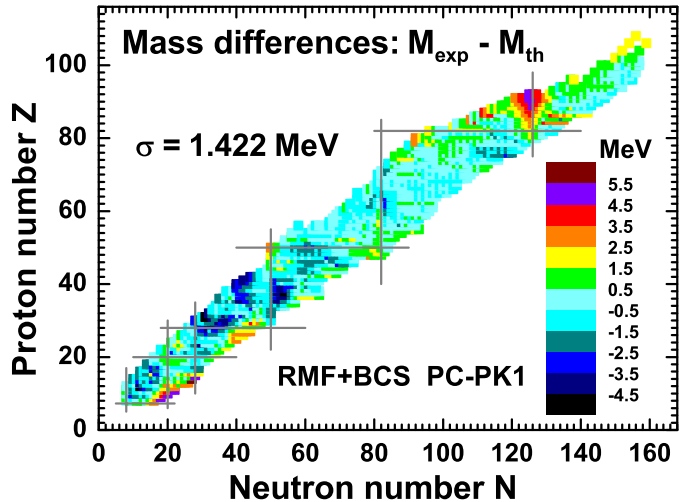}
  \caption{Mass differences between the experimental data~\cite{Audi2003Nucl.Phys.A337} and the covariant density functional theory calculations with the point-coupling interaction PC-PK1~\cite{Zhao2010Phys.Rev.C54319}.}
  \label{fig2-1}
  \end{figure}

  By using the PC-PK1, the masses of nuclei with neutron number $N\ge8$ and proton number $Z\ge8$ in the atomic mass evaluation of 2003 (AME03)~\cite{Audi2003Nucl.Phys.A337} have been calculated in the framework CDFT with the axial symmetry. The mass differences between the experimental data and the calculated results are shown in Fig.~\ref{fig2-1}.
  It is found that the CDFT with PC-PK1 can reproduce the experimental data quite well and the corresponding root-mean-square (rms) deviation is 1.422 MeV, which is much smaller than the rms deviation value of 2.25 MeV given by the meson-exchange effective interaction TMA~\cite{Geng2005Prog.Theor.Phys.785}. Note that in the results given by PC-PK1, the rotational correction energy due to the violation of the rotational symmetry is considered for deformed nuclei with the moment of inertia ${\cal I}>3.5~\hbar^2/{\rm MeV}$ by $E_{\rm rot}=-\frac{\hbar^2}{2b\cal I}\langle \hat{J}^2\rangle$, where $\hat{J}$ is the angular momentum operator and $\cal I$ is the moment of inertia calculated from the Inglis-Belyaev formula~\cite{Inglis1956Phys.Rev.1786,Belyaev1961Nucl.Phys.A322,Volkov1972Phys.Lett.B1}. The parameter $b$ is chosen as 1, 1.4, and 1.6 for even-even, odd-A, and odd-odd nuclei, respectively. As discussed in Ref.~\cite{Zhao2012Phys.Rev.C64324},  further improvements on a proper treatment of the rotational correction energy for odd nuclei is necessary.

 \subsection{Tilted axis cranking CDFT}\label{Subsec2.3}

In order to describe the magnetic rotations in a microscopical and self-consistent way,
the tilted axis cranking CDFT should be developed. This approach has been realized based on either the meson-exchange interaction~\cite{Madokoro2000Phys.Rev.C61301,Peng2008Phys.Rev.C24313} or the point-coupling interaction~\cite{Zhao2011Phys.Lett.B181}. The TAC-RMF approach based on the meson-exchange interaction is described in detail in Ref.~\cite{Peng2008Phys.Rev.C24313}. In the following, the main discussion will be focused on the TAC-RMF approach based on the point-coupling interaction~\cite{Zhao2011Phys.Lett.B181}.

Assuming that the nucleus rotates around an axis in the $xz$ plane, the Lagrangian in Eq.~(\ref{EQ:LAG}) is transformed into a frame rotating uniformly with a constant rotational frequency,
\begin{equation}
  \bm{\Omega}=(\Omega_x,0,\Omega_z)=(\Omega\cos\theta_\Omega,0,\Omega\sin\theta_\Omega),
\end{equation}
where $\theta_\Omega:=\sphericalangle(\bm{\Omega},\bm{e}_x)$ is the tilted angle between the cranking axis and the $x$ axis. From this rotating Lagrangian, the equation of motion for the nucleons can be derived equivalently by either starting from a special relativistic transformation~\cite{Koepf1989Nucl.Phys.A61}, or more generally adopting the tetrad formalism
in the framework of general relativistic theory~\cite{Madokoro1997Phys.Rev.C2934}. One can thus finds
 \begin{equation}\label{Eq.Dirac}
   [\bm{\alpha}\cdot(-i\bm{\nabla}-\bm{V})+\beta(m+S)
    +V-\bm{\Omega}\cdot\hat{\bm{J}}]\psi_k=\epsilon_k\psi_k,
 \end{equation}
where $\hat{\bm{J}}=\hat{\bm{L}}+\frac{1}{2}\hat{\bm{\Sigma}}$ is the total angular momentum of the nucleon spinors, and $\epsilon_k$ represents the single-particle Routhians for nucleons. The relativistic fields $S(\bm{r})$ and $V^\mu(\bm{r})$ read
\begin{subequations}\label{Eq.poten}
 \begin{eqnarray}
   S(\bm{r})&=&\alpha_S\rho_S+\beta_S\rho_S^2+\gamma_S\rho_S^3+\delta_S\triangle\rho_S, \\
   V(\bm{r})&=&\alpha_V\rho_V+\gamma_V\rho_V^3+\delta_V\triangle \rho_V+\tau_3\alpha_{TV} \rho_{TV}+\tau_3\delta_{TV}\triangle \rho_{TV}+eA^0, \\
   \bm{V}(\bm{r})&=&\alpha_V \bm{j}_V+\gamma_V(\bm{j}_V)^3+\delta_V\triangle \bm{j}_V+\tau_3\alpha_{TV} \bm{j}_{TV}+\tau_3\delta_{TV}\triangle \bm{j}_{TV}+e\bm{A}.
 \end{eqnarray}
\end{subequations}
As usual, it is assumed that the nucleon single-particle states do not mix isospin, i.e., the single-particle states are eigenstates of $\tau_3$. Therefore only the third component of isovector potentials survives. The Coulomb field $A^0(\bm{r})$ is determined by Poisson's equation
\begin{equation}
  -\triangle A^0(\bm{r}) = e\rho_c.
\label{Laplace}
\end{equation}
The spatial components of the electromagnetic vector potential $\bm{A}(\bm{r})$ are neglected since their contributions are extremely small~\cite{Koepf1989Nucl.Phys.A61,Koepf1990Nucl.Phys.A279}.

Since the Coriolis term $\bm{\Omega}\cdot\hat{\bm{J}}$ in the Dirac equation~(\ref{Eq.Dirac}) breaks time reversal symmetry in the intrinsic frame, the nucleon currents are induced and as a consequence the spatial components of the vector potential $\bm{V}(\bm{r})$. The densities and currents in Eqs.~(\ref{Eq.poten}) have the form
\begin{subequations}
 \begin{eqnarray}\label{crank:currents}
   \rho_S(\bm{r})     &=& \sum_{i=1}^A \bar\psi_i(\bm{r})\psi_i(\bm{r})        , \\
   \rho_V(\bm{r})     &=& \sum_{i=1}^A \psi_i^\dagger(\bm{r})\psi_i(\bm{r})        , \\
   \bm{j}_V(\bm{r})   &=& \sum_{i=1}^A \psi_i^\dagger(\bm{r})\bm{\alpha}\psi_i(\bm{r}) , \\
   \rho_{TV}(\bm{r})  &=& \sum_{i=1}^A \psi_i^\dagger(\bm{r})\tau_3\psi_i(\bm{r})      , \\
   \bm{j}_{TV}(\bm{r})&=& \sum_{i=1}^A \psi_i^\dagger(\bm{r})\bm{\alpha}\tau_3\psi_i(\bm{r})    , \\
   \rho_{c}(\bm{r})   &=& \sum_{i=1}^A \psi_i^\dagger(\bm{r})\frac{1-\tau_3}{2}\psi_i(\bm{r}).
 \end{eqnarray}
\end{subequations}
Here, the ``no-sea'' approximation is also adopted, i.e., the sums run over only the particles states in the Fermi sea and the contribution of the negative-energy states are neglected.

By solving the equation of motion iteratively, one finally obtains the total energy in the laboratory frame
\begin{equation}
  \label{Eq.Etot}
   E_{\rm tot} = E_{\rm kin} + E_{\rm int} + E_{\rm cou} + E_{\rm c.m.},
\end{equation}
which is composed of a kinetic part
\begin{equation}
   E_{\rm kin} = \int d^3\bm{r}\sum\limits_{i=1}^A\psi_i^\dagger[\bm{\alpha}\cdot\bm{p}+\beta m]\psi_i,
\end{equation}
an interaction part
\begin{eqnarray}
    E_{\rm int} &=&\int d^3\bm{r} \left\{\frac{1}{2}\alpha_S\rho_S^2+\frac{1}{3}\beta_S\rho_S^3+\frac{1}{4}\gamma_S\rho_S^4
    +\frac{1}{2}\delta_S\rho_S\Delta\rho_S\right.\nonumber \\
    &&+\frac{1}{2}\alpha_V(\rho_V^2-\bm{j}\cdot\bm{j})
    +\frac{1}{2}\alpha_{TV}(\rho_{TV}^2-\bm{j}_{TV}\cdot\bm{j}_{TV})\nonumber \\
    &&+\frac{1}{4}\gamma_V(\rho_V^2-\bm{j}\cdot\bm{j})^2
    +\frac{1}{2}\delta_V(\rho_V\Delta\rho_V-\bm{j}\Delta\bm{j})\nonumber \\
    &&+\left.\frac{1}{2}\delta_{TV}(\rho_{TV}\Delta\rho_{TV}-\bm{j}_{TV}\Delta\bm{j}_{TV})\right\},
\end{eqnarray}
an electromagnetic part
\begin{equation}
    E_{\rm cou} = \int d^3\bm{r}\frac{1}{2}eA_0\rho_c,
\end{equation}
and the center-of-mass (c.m.) correction energy $E_{\rm c.m.}$ accounting for the treatment of center-of-mass motion.

For each rotational frequency $\Omega$, the expectation values of the angular momentum components $\bm{J}=(J_x,J_y,J_z)$
in the intrinsic frame are given by
\begin{subequations}
\begin{eqnarray}
    J_x&=&\langle\hat{J}_x\rangle=\sum\limits_{i=1}^Aj^{(i)}_x, \\
    J_y&=& 0, \\
    J_z&=&\langle\hat{J}_z\rangle=\sum\limits_{i=1}^Aj^{(i)}_z,
\end{eqnarray}
\end{subequations}
and by means of the semiclassical cranking condition
\begin{equation}
 J = \sqrt{\langle\hat{J}_x\rangle^2+\langle\hat{J}_z\rangle^2}\equiv\sqrt{I(I+1)},
\end{equation}
one can relate the rotational frequency $\Omega$ to the angular momentum quantum number $I$ in the rotational band.

The orientation of the angular momentum vector $\bm{J}$ is represented by the angle $\theta_J:=\sphericalangle(\bm{J},\bm{e}_x)$ between the angular momentum vector $\bm{J}$ and the $x$ axis. As mentioned before, in a fully self-consistent calculation, the orientation $\theta_J$ of the angular momentum $\bm{J}$ should be identical to the orientation  $\theta_\Omega$ of the angular velocity $\bm{\Omega}$.

The quadrupole moments $Q_{20}$ and $Q_{22}$ are calculated by
\begin{subequations}
\begin{eqnarray}
    Q_{20}&=&\sqrt{\frac{5}{16\pi}}\langle3z^2-r^2\rangle, \\
    Q_{22}&=&\sqrt{\frac{15}{32\pi}}\langle x^2-y^2\rangle,
\end{eqnarray}
\end{subequations}
and the deformation parameters $\beta$ and $\gamma$ can thus be extracted from
\begin{subequations}
\begin{eqnarray}
    \beta&=&\sqrt{a_{20}^2+2a_{22}^2}, \\
   \gamma&=&\arctan\left[\sqrt{2}\frac{a_{22}}{a_{20}}\right],
\end{eqnarray}
\end{subequations}
by using the relations
\begin{subequations}
\begin{eqnarray}
    Q_{20}&=&\frac{3A}{4\pi}R_0^2a_{20}, \\
    Q_{22}&=&\frac{3A}{4\pi}R_0^2a_{22},
\end{eqnarray}
\end{subequations}
with $R_0 = 1.2A^{1/3}~\rm fm$. Note that the sign convention in Ref.~\cite{Ring1980} is adopted for the definition of $\gamma$ here.

The nuclear magnetic moment, in units of the nuclear magneton, is given by
\begin{equation}
   \bm{\mu} = \sum\limits_{i=1}^A\int d^3r\left[\frac{mc^2}{\hbar c}q\psi^\dagger_i(\bm{r})\bm{r}\times\bm{\alpha}\psi_i(\bm{r})+\kappa\psi^\dagger_i(\bm{r})\beta\bm{\Sigma}\psi_i(\bm{r})\right],
\end{equation}
where the charge $q$ ($q_p=1$ for protons and $q_n=0$ for neutrons) is in units of $e$, $m$ the nucleon mass, and $\kappa$ the free anomalous gyromagnetic ratio of the nucleon ($\kappa_p=1.793$ and $\kappa_n=-1.913$).

From the quadrupole moments and the magnetic moment, the $B(M1)$ and $B(E2)$ transition probabilities can be derived in semiclassical approximation
\begin{subequations}
\begin{eqnarray}
   B(M1) &=& \frac{3}{8\pi}\mu_{\bot}^2 =\frac{3}{8\pi}(\mu_x\sin\theta_J-\mu_z\cos\theta_J)^2,\\
   B(E2) &=& \frac{3}{8}\left[ Q^p_{20}\cos^2\theta_J+\sqrt{\frac{2}{3}}Q^p_{22}(1+\sin^2\theta_J)\right]^2,
\end{eqnarray}
\end{subequations}
where $Q^p_{20}$ and $Q^p_{22}$ corresponds to the quadrupole moments of protons.

\subsection{Numerical techniques}\label{Subsec2.4}

{\bf Orientation constraint}
In the usual PAC programs (one-dimensional cranking), the principal axes of the densities and fields are implemented to be along the $x$, $y$, and $z$ axis. For the TAC code (two-dimensional cranking), it allows for arbitrary rotations of the density distributions around the intrinsic $y$ axis. The freedom of rotations around the $y$ axis can lead (in particular for $\Omega=0$ and for small $\Omega$ values) to instabilities during the iterative solution because the solutions with different orientations in the $xz$ plane are degenerate. Therefore, the $x$, $y$, and $z$ axes are enforced to be identical with the principal axes of the density distribution by introducing a quadratic constraint~\cite{Ring1980} for the expectation value of the quadrupole moment
\begin{equation}
\langle{Q_{2-1}}\rangle=-\sqrt{\frac{15}{8\pi}}\left\langle
xz\right\rangle =0, \label{E12}%
\end{equation}
i.e., by minimizing
\begin{equation}
\left\langle H^{\prime}\right\rangle =\left\langle H\right\rangle +\frac{1}{2}C\left( \langle{{Q}}_{2-1}\rangle-a_{2-1}\right)  ^{2},
\label{for:constrain2}
\end{equation}
with $a_{2-1}=0$, and $C$ being a spring constant, which, if properly
chosen, has no influence on the final result.

{\bf Expansion in harmonic oscillator basis}
In the code for the solution of the relativistic tilted axis cranking equations, the Dirac spinors are expanded in terms of three-dimensional harmonic oscillator wave functions in Cartesian coordinates,
\begin{equation}
\varphi_{n_{x}n_{y}n_{z}}({\bm{r}})=\langle{\bm{r}}|n_{x},n_{y},n_{z}\rangle=\varphi_{n_{x}}(x)\varphi_{n_{y}}(y)\varphi_{n_{z}}(z).
\label{basis-spinors}
\end{equation}
The normalized oscillator function $\varphi_{n_{k}}(x_{k})$ in $k$-direction ($x_{k}=$ $x,y,z$) are given by
\begin{equation}
\displaystyle\varphi_{n_{k}}(x_{k})=\frac{N_{n_{k}}}{\sqrt{b_{k}}}H_{n_{k}}(\frac{x_{k}}{b_{k}})\exp[{-\frac{1}{2}(\frac{x_{k}}{b_{k}})^{2}}],
\label{E18}
\end{equation}
where $N_{n}=(\sqrt{\pi}2^{n}n!)^{-1/2}$ is a normalization factor and
\begin{equation}
H_{n}(\xi)=(-1)^{n}e^{\xi^{2}}\frac{d^{n}}{d\xi^{n}}e^{-\xi^{2}}%
\end{equation}
are the Hermite polynomials~\cite{Abramowitz1965}.

We can take the following basis states
\begin{equation}
\varphi_{\alpha}({\mathbf{r}},s)=\langle{\mathbf{r}},s|\alpha\rangle=i^{n_{y}%
}\varphi_{n_{x}}(x)\varphi_{n_{y}}(y)\varphi_{n_{z}}(z)\frac{1}{\sqrt{2}%
}\left(
\begin{array}
[c]{c}%
1\\
(-1)^{n_{x}+1}%
\end{array}
\right)  , \label{basis1}%
\end{equation}
and
\begin{equation}
\varphi_{\overline{\alpha}}({\mathbf{r}},s)=\langle{\mathbf{r}},s|\overline
{\alpha}\rangle=(-i)^{n_{y}}\varphi_{n_{x}}(x)\varphi_{n_{y}}(y)\varphi
_{n_{z}}(z)\frac{1}{\sqrt{2}}\left(
\begin{array}
[c]{c}%
(-1)^{n_{x}+1}\\
-1
\end{array}
\right)  , \label{basis2}%
\end{equation}
which correspond to the eigenfunctions of the simplex operation with the positive ($+i$) and negative ($-i$) eigenvalues, respectively.
The phase factor $i^{n_{y}}$ has been added in order to have real matrix elements for the Dirac equation~\cite{Girod1983Phys.Rev.C2317}.

The Dirac spinor for the nucleon has the form
\begin{equation}
\psi_{i}({\mathbf{r}})=\left(
\begin{array}
[c]{l}%
f_{i}({\mathbf{r}},s)\\
ig_{i}({\mathbf{r}},s)
\end{array}
\right)  \chi_{i}(t), \label{ctrmf:WaveFunction}%
\end{equation}
where $\chi_{i}(t)$ is the isospin part. In the tilted axis cranking calculations, the simplex symmetry is violated, and therefore the large and small components of the wave function in Eq.~(\ref{ctrmf:WaveFunction}) have to be written as linear combinations of the sets (\ref{basis1}) and (\ref{basis2}) with different simplex:
\begin{equation}%
\begin{array}
[c]{ccc}%
f_{i}({\mathbf{r}},s) & = & \sum\limits_{\alpha}f_{\alpha i}|\alpha
\rangle+\sum\limits_{\bar{\alpha}}f_{\bar{\alpha}i}|\bar{\alpha}\rangle,\\
g_{i}({\mathbf{r}},s) & = & \sum\limits_{\tilde{\alpha}}g_{\tilde{\alpha}%
i}|\tilde{\alpha}\rangle+\sum\limits_{\overline{\tilde{\alpha}}}%
g_{\overline{\tilde{\alpha}}i}|\overline{\tilde{\alpha}}\rangle.
\end{array}
\label{expansion}%
\end{equation}
Since the large and small components in the Dirac equation have different
parity, the sums in the expansions for the large and the small components have
to run over oscillator quantum numbers with even $\,N=n_{x}+n_{y}+n_{z}$ or
odd $N$ respectively. This is indicated in Eq.~(\ref{expansion}) by the
indices $\alpha$ and $\tilde{\alpha}$.

On this basis, the solution of Dirac equation (\ref{Eq.Dirac}) is obtained by
the matrix diagonalization
\begin{equation}
\mathcal{H}\left(
\begin{array}
[c]{c}%
f_{\alpha i}\\
f_{\overline{\alpha}i}\\
g_{\tilde{\alpha}i}\\
g_{\overline{\tilde{\alpha}}i}%
\end{array}
\right)  =\varepsilon_{i}\left(
\begin{array}
[c]{c}%
f_{\alpha i}\\
f_{\overline{\alpha}i}\\
g_{\tilde{\alpha}i}\\
g_{\overline{\tilde{\alpha}}i}%
\end{array}
\right)  ,
\end{equation}
where the Hamiltonian matrix $\mathcal{H}$ has the form
%{\tiny
{\small
\begin{equation}
\left(
\begin{array}
[c]{cccc}%
{\left\langle {\alpha}\right\vert }{M^{\ast}+V-\mbox{\boldmath$\Omega$}\hat
{\bm{J}}}{\left\vert \alpha^{\prime}\right\rangle } & {\left\langle
{\alpha}\right\vert }{M^{\ast}+V-\mbox{\boldmath$\Omega$}\hat{\bm{J}}%
}{\left\vert \overline{\alpha}^{\prime}\right\rangle } & {\left\langle
{\alpha}\right\vert }%
{\mbox{\boldmath$\sigma$}(\mbox{\boldmath$\nabla$}-i{\bm{V}})}{\left\vert
\tilde{\alpha}^{\prime}\right\rangle } & {\left\langle {\alpha}\right\vert
}{\mbox{\boldmath$\sigma$}(\mbox{\boldmath$\nabla$}-i{\bm{V}})}{\left\vert
\overline{\tilde{\alpha}}^{\prime}\right\rangle }\\
{\left\langle {\overline{\alpha}}\right\vert }{M^{\ast}%
+V-\mbox{\boldmath$\Omega$}\hat{\bm{J}}}{\left\vert \alpha^{\prime
}\right\rangle } & {\left\langle {\overline{\alpha}}\right\vert }{M^{\ast
}+V-\mbox{\boldmath$\Omega$}\hat{\bm{J}}}{\left\vert \overline{\alpha
}^{\prime}\right\rangle } & {\left\langle {\overline{\alpha}}\right\vert
}{\mbox{\boldmath$\sigma$}(\mbox{\boldmath$\nabla$}-i{\bm{V}})\left\vert
\tilde{\alpha}^{\prime}\right\rangle } & {\left\langle {\overline{\alpha}%
}\right\vert }{\mbox{\boldmath$\sigma$}(\mbox{\boldmath$\nabla$}-i{\bm{V}%
})}{\left\vert \overline{\tilde{\alpha}}^{\prime}\right\rangle }\\
{\left\langle \tilde{\alpha}\right\vert }%
{-\mbox{\boldmath$\sigma$}(\mbox{\boldmath$\nabla$}-i{\bm{V}})}{\left\vert
\alpha^{\prime}\right\rangle } & {\left\langle \tilde{\alpha}\right\vert
}{-\mbox{\boldmath$\sigma$}(\mbox{\boldmath$\nabla$}-i{\bm{V}}%
)}{\left\vert \overline{\alpha}^{\prime}\right\rangle } & {\left\langle
\tilde{\alpha}\right\vert }{-M^{\ast}+V-\mbox{\boldmath$\Omega$}\hat
{\bm{J}}}{\left\vert \tilde{\alpha}^{\prime}\right\rangle } &
{\left\langle \tilde{\alpha}\right\vert }{-M^{\ast}%
+V-\mbox{\boldmath$\Omega$}\hat{\bm{J}}}{\left\vert \overline
{\tilde{\alpha}}^{\prime}\right\rangle }\\
{\left\langle \overline{\tilde{\alpha}}\right\vert }%
{-\mbox{\boldmath$\sigma$}(\mbox{\boldmath$\nabla$}-i{\bm{V}})}{\left\vert
\alpha^{\prime}\right\rangle } & {\left\langle \overline{\tilde{\alpha}%
}\right\vert }{-\mbox{\boldmath$\sigma$}(\mbox{\boldmath$\nabla$}-i{\bm{V}%
})}{\left\vert \overline{\alpha}^{\prime}\right\rangle } & {\left\langle
\overline{\tilde{\alpha}}\right\vert }{-M^{\ast}%
+V-\mbox{\boldmath$\Omega$}\hat{\bm{J}}}{\left\vert \tilde{\alpha}%
^{\prime}\right\rangle } & {\left\langle \overline{\tilde{\alpha}}\right\vert
}{-M^{\ast}+V-\mbox{\boldmath$\Omega$}\hat{\bm{J}}}{\left\vert
\overline{\tilde{\alpha}}^{\prime}\right\rangle }%
\end{array}
\right)  .
\end{equation}
}
Note that the Coriolis term ${\mbox{\boldmath$\Omega$}\hat{\bm{J}}}$ breaks the invariance with respect to time reversal and with respect to rotations around the $x$ axis as well as around the $z$ axis. Therefore, only the invariance of space reflection $\mathcal{P}$ and the combination of time reversal and reflection in $y$ direction $\mathcal{P}_{y}\mathcal{T}$ are valid and used in the code.

For the evaluation of the Coulomb field, due to its long range character, an expansion in harmonic oscillator states is very difficult and therefore the standard Green function method~\cite{Vautherin1973Phys.Rev.C296} is used for the calculation of the Coulomb field in each step of the iteration.

{\bf Configuration constraint}
Normally the rotation bands are built on specific proton and neutron configurations. The orbits to be blocked are usually given in the spherical basis, i.e., by the spherical quantum numbers $\left\vert nljm\right\rangle $. The equations of motion are solved by expanding the Dirac spinors in terms of the three-dimensional harmonic oscillator functions in the Cartesian basis Eq.~(\ref{expansion}) labeled by the quantum number $\left\vert {n_{x},n_{y},n_{z},\pm i}\right\rangle $. In order to identify which orbits have to be blocked, one need to transform the wave functions from the Cartesian basis with the quantum number $\left\vert {n_{x}n_{y}n_{z}\pm i}\right\rangle$ to a spherical basis with the quantum numbers $\left\vert {nljm}\right\rangle$ using the techniques given in Refs.~\cite{Chasman1967Nucl.Phys.A401,Talman1970Nucl.Phys.A273}. Consequently, one can block the levels which have the maximal overlap with the required $\left\vert {nljm}\right\rangle $ orbits. These techniques are considerably simplified, if we work in an isotropic Cartesian basis with identical basis parameters $b_{x}=b_{y}=b_{z}=b$ in Eq.~(\ref{E18}).

To describe a rotational band, one should keep the corresponding configuration fixed for a set of increasing values of rotational frequency $\Omega$. Because of the considerable $K$-mixing in TAC solutions and the high level density in the 3-dimensional calculations, it may occur that the configuration is changing with the iteration and also the different rotational frequency. To constrain the specific configuration we are interested in, one can adopt the following prescription: starting from the Dirac level $|\psi_{i}(\Omega_{n})\rangle$ blocked for $\Omega=\Omega_{n}$, one can block for $\Omega=\Omega_{n+1}$ the level $|\psi_{j}(\Omega_{n+1})\rangle$ which maximizes the overlap $\langle \psi_{i}(\Omega_{n})|\psi_{j}(\Omega_{n+1})\rangle$, i.e.,
\begin{equation}
\left\langle {\psi_{j}(\Omega+\delta\Omega)}|\psi_{i}(\Omega)\right\rangle
=1+{\mathcal{O}}(\delta\Omega). \label{par-tra}%
\end{equation}
For infinitesimal step sizes, this condition corresponds to the so-called parallel transport~\cite{Simon1983Phys.Rev.Lett.2167,Bengtsson1989Nucl.Phys.A56}.

\section{Celebrated Magnetic rotation in $^{198}$Pb}

\subsection{Brief historical overview}

The observation of the $\Delta=1$ rotational-like structures in neutron deficient Pb nuclei in the early 1990s opened a new era for magnetic rotation~\cite{Clark2000Annu.Rev.Nucl.Part.Sci.1,Frauendorf2001Rev.Mod.Phys.463,Hubel2005Prog.Part.Nucl.Phys.1}.
Long cascades of M1 transitions were firstly observed in Pb nuclei in the early 1990's~\cite{Clark1992Phys.Lett.B247,Baldsiefen1992Phys.Lett.B252, Kuhnert1992Phys.Rev.C133,Clark1993Nucl.Phys.A562,Baldsiefen1994Nucl.Phys.A574}. With improved detector techniques and lots of efforts, in 1997 the lifetime measurements based on the Doppler-shift attenuation method (DSAM) for four M1-bands in the nuclei $^{198}\rm Pb$ and $^{199}\rm Pb$  provided a clear evidence for magnetic rotation~\cite{Clark1997Phys.Rev.Lett.1868}. Subsequently, another  experiment using the recoil distance method (RDM) in $^{198}\rm Pb$ together with the results of the DSAM experiment provided further support to the shears mechanism~\cite{Krucken1998Phys.Rev.C58}.
Many works along this line for almost 10 years have been devoted to
the magnetic rotation in Pb isotopes which becomes a classic example and
has received wide attention from theoretical and experimental aspects.
Naturally, it is a good test ground for various theory aiming at the description of the MR.

In the framework of the pairing plus quadrupole model, the tilted axis cranking approach~\cite{Frauendorf1993Nucl.Phys.A259,Frauendorf2000Nucl.Phys.A115,Frauendorf2001Rev.Mod.Phys.463}
has been used to describe the magnetic rotation from the very beginning and reproduce the experimental reduced $B(M1)$ values of magnetic dipole bands in $^{198,199}$Pb very well~\cite{Clark1997Phys.Rev.Lett.1868}.
In addition, the shell model~\cite{Frauendorf1996Nucl.Phys.A601} and the many-particles-plus-rotor model~\cite{Carlsson2006Phys.Rev.C74} have also been applied to investigate magnetic rotation in the Pb region.

With its many success in describing nuclear phenomena in
stable as well as in exotic nuclei~\cite{Ring1996Prog.Part.Nucl.Phys.193,Vretenar2005Phys.Rep.101,Meng2006Prog.Part.Nucl.Phys.470}, the CDFT has been generalized to the cranking CDFT~\cite{Koepf1990Nucl.Phys.A279} and the tilted axis cranking CDFT~\cite{Madokoro2000Phys.Rev.C61301,Peng2008Phys.Rev.C24313,Zhao2011Phys.Lett.B181}.
In particular, the newly developed tilted axis cranking covariant density functional theory (TAC-CDFT) based on point-coupling interactions~\cite{Zhao2011Phys.Lett.B181}
includes significant improvements by implanting the simplex symmetry and reduces computation time, which makes it possible to investigate the MR in the heaviest mass region observed so far.

As the magnetic rotation in Pb isotopes
is typical and famous examples, in the following the band 1 in $^{198}$Pb~\cite{Clark1993Nucl.Phys.A562} will be chosen as an example to demonstrate the relativistic self-consistent
description of the MR characteristics. The calculations~\cite{Yu2012Phys.Rev.C24318} have been carried out with the covariant point-coupling density functional PC-PK1~\cite{Zhao2010Phys.Rev.C54319}.
The detailed and numerical techniques can be
found in Ref.~\cite{Yu2012Phys.Rev.C24318} and references therein.

\subsection{Single particle Routhian and configuration}

\begin{figure*}
\centerline{
\includegraphics[width=10cm]{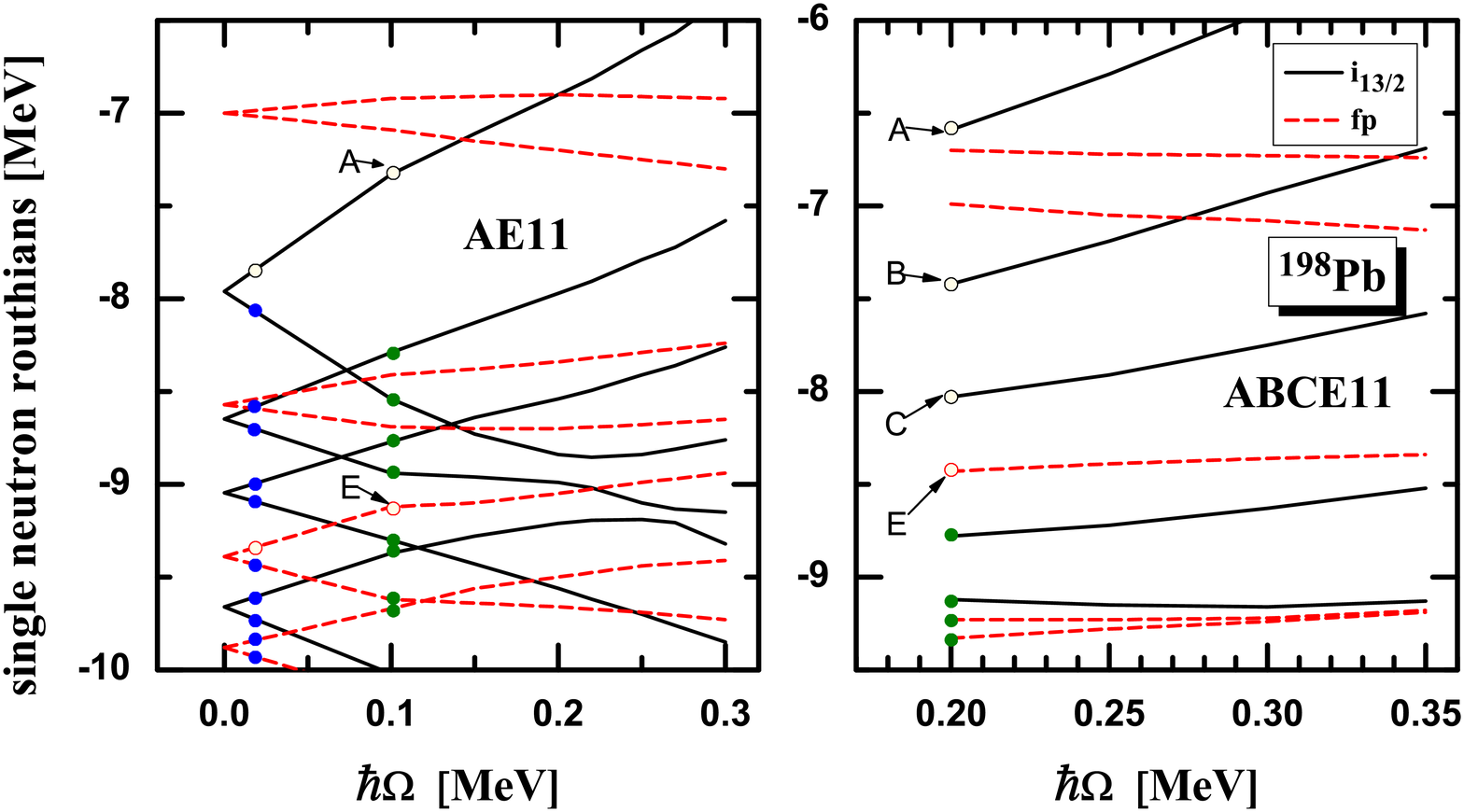}
}\caption{(Color online) Single particle routhians for the neutrons in $^{198}$Pb as a function of the rotational frequency based on the configurations AE11 and ABCE11. The blue dots indicate the occupied levels at $\Omega=0$ and the green dots indicate the occupied levels at the band heads with the configuration AE11 (left panel) and ABCE11 (right panel). Further details are given in the text. Taken from Ref.~\cite{Yu2012Phys.Rev.C24318}.}
\label{fig3-1}
\end{figure*}

In Ref.~\cite{Yu2012Phys.Rev.C24318}, TAC based CDFT calculation has been performed with the proton configuration $\pi[s_{1/2}^{-2}h_{9/2}i_{13/2}]11^-$ suggested as in Ref.~\cite{Clark1993Nucl.Phys.A562}. The single particle Routhians for the neutrons in the nucleus $^{198}$Pb are respectively shown as a function of the rotational frequency $\Omega$ for the two configurations AE11 and ABCE11 in Fig.~\ref{fig3-1}. The positive parity levels belonging to the $\nu i_{13/2}$ orbit are given by full black curves and the neutron levels with negative parity $(pf)$ are indicated by dashed red curves. In band 1, a backbending phenomenon has been observed caused by the alignment of a pair of $i_{13/2}$ neutrons.
Before the backbending, the neutron configurations $\nu [i_{13/2}^{-1}(fp)^{-1}]$ has been assigned to band 1 in $^{198}\rm Pb$. After the backbending, it  becomes as $\nu[ i_{13/2}^{-3}(fp)^{-1}]$~\cite{Amita2000At.DataNucl.DataTables283,Gorgen2001Nucl.Phys.A683}. As in Ref.~\cite{Baldsiefen1994Nucl.Phys.A574}, a short hand notation is used for these configurations: A, B, C and D denote $\nu i_{13/2}$ holes with positive parity and by E denotes a neutron hole with negative parity (mainly of $f_{5/2}$ and $p_{3/2}$ origin). The proton configuration $\pi[s_{1/2}^{-2}h_{9/2}i_{13/2}]11^-$ is abbreviated by its spin number 11. Therefore the neutron configurations $\nu [i_{13/2}^{-1}(fp)^{-1}]$ is referred as AE11 before the backbending, and $\nu[ i_{13/2}^{-3}(fp)^{-1}]$ as ABCE11 after the backbending.

%%%%%%%%%%%%%%%%%%%%%%%%%%%%%%%%%%%%%%%%%%%%%%%%%%%%%%%%%%%%%%%%%%%%%%%%%%%%%%%%%%%%%%%%%%
%%%%%%%%%%%%%%%%%%%%%%%%%%%%%%%%%%%%%%%%%%%%%%%%%%%%%%%%%%%%%%%%%%%%%%%%%%%%%%%%%%%%%%%%%%
\subsection{Energy spectra}

%%%%%%%%%%%%%%%%%%%%%%%%%%%%%%%%%%%%%%%%%%%%%%%%%%%%%%%%%%%%%%%%%%%%%%%%%%%%%%%%%%%%%%%%%%
\begin{figure*}
\centerline{
\includegraphics[width=8cm]{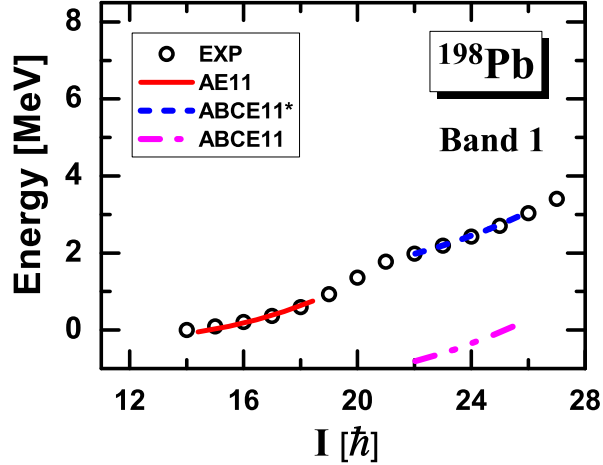}
} \caption{(Color online)
Energy spectra in the TAC-CDFT calculations compared with the data~\cite{Gorgen2001Nucl.Phys.A683} for band 1 in $^{198}\rm Pb$. The energies at $I= 15\hbar$ is taken as references for the band 1 in $^{198}\rm Pb$. Energies for the configurations ABCE11* in $^{198}\rm Pb$ are renormalized to the energies at $I=22\hbar$. Taken from Ref.~\cite{Yu2012Phys.Rev.C24318}.}
\label{fig3-2}
\end{figure*}
%%%%%%%%%%%%%%%%%%%%%%%%%%%%%%%%%%%%%%%%%%%%%%%%%%%%%%%%%%%%%%%%%%%%%%%%%%%%%%%%%%%%%%%%%%

The calculated energy spectra of the band 1 in $^{198}\rm Pb$ are show in comparison with the data~\cite{Gorgen2001Nucl.Phys.A683} in Fig.~\ref{fig3-2}. For certain regions of angular momenta, the calculated values are missing, as for instance, $I=19-21\hbar$ in band 1 in $^{198}\rm Pb$.
 As discussed in Ref.~\cite{Peng2008Phys.Rev.C24313}, it is due to the level crossing connected with the backbending phenomenon, and no converged solutions could be found for these angular momentum values. It can be seen that the TAC-CDFT calculations reproduce well the experimental energies but underestimate the particle-hole excitation energies at the band head of the configurations ABCE11 in $^{198}\rm Pb$. In comparison with the PQTAC calculations~\cite{Frauendorf1993Nucl.Phys.A259,Chmel2007Phys.Rev.C75}, these underestimations can be explained by the pairing correlations and will be further investigated in the future. At the moment, these underestimations are compensated by choosing different references for the configurations involved.

%%%%%%%%%%%%%%%%%%%%%%%%%%%%%%%%%%%%%%%%%%%%%%%%%%%%%%%%%%%%%%%%%%%%%%%%%%%%%%%%%%%%%%%%%%%%
\begin{figure*}
\centerline{
\includegraphics[width=8cm]{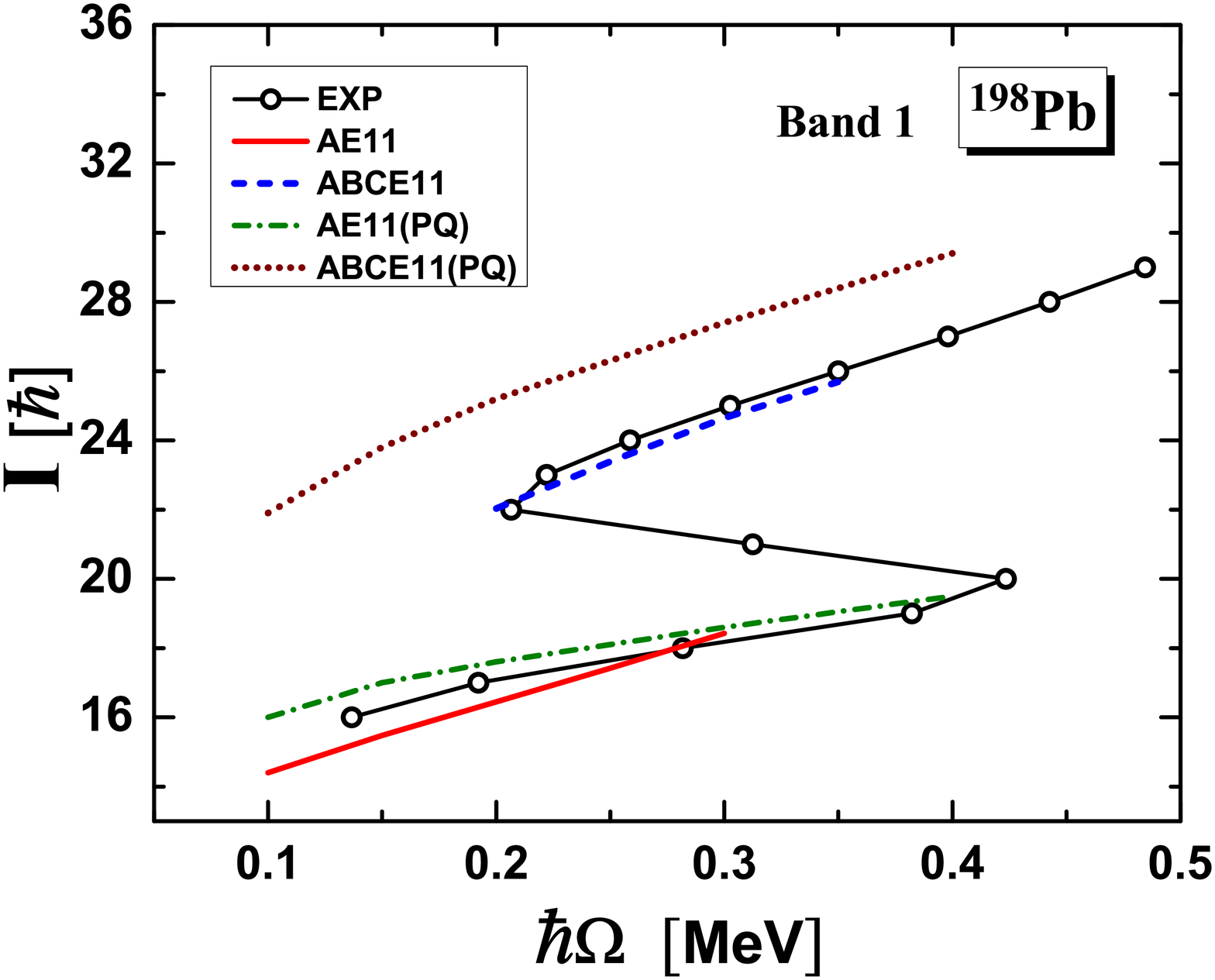}
} \caption{(Color online) Angular momenta as  functions of the
rotational frequency in the TAC-CDFT calculations
compared with the data~\cite{Gorgen2001Nucl.Phys.A683} and the PQTAC results~\cite{Chmel2007Phys.Rev.C75} for band 1 in $^{198}\rm Pb$.
The configurations with ``(PQ)'' denote the corresponding
results of PQTAC calculations. Taken from Ref.~\cite{Yu2012Phys.Rev.C24318}.}
\label{fig3-3}
\end{figure*}
%%%%%%%%%%%%%%%%%%%%%%%%%%%%%%%%%%%%%%%%%%%%%%%%%%%%%%%%%%%%%%%%%%%%%%%%%%%%%%%%%%%%%%%%%%%%

The experimental rotational frequency $\Omega_{\rm exp}$ is extracted from the energy spectra by the relation
 \begin{equation}
    \hbar\mathit\Omega_{\rm exp}\approx\frac{dE}{dI}=\frac{1}{2}[E_\gamma(I+1\rightarrow I)+E_\gamma(I\rightarrow I-1)].
 \end{equation}
 In Fig.~\ref{fig3-3}, the calculated total angular momenta of the band 1 in $^{198}\rm Pb$ as functions of the rotational frequency are shown in comparison with the experimental data~\cite{Gorgen2001Nucl.Phys.A683} and the PQTAC results~\cite{Chmel2007Phys.Rev.C75}.
 It is found that both the TAC-CDFT and the PQTAC results agree well with the experimental data. This shows that the TAC calculations can reproduce the relative changes of the moment of inertia within the different bands rather well.
 The TAC calculations support that the backbendings arise through an excitation of a neutron-hole pair in the $i_{13/2}$ shell, i.e. by the transitions in the configurations AE11$\rightarrow$ABCE11 in band 1 of $^{198}\rm Pb$. In detail, before the backbending the spins values found in the TAC-CDFT and PQTAC models differ from experimental values up to $2\hbar$. After the backbending, the PQTAC result for the band 1 in $^{198}\rm Pb$ is nearly $3\hbar$ larger than the experimental values and the TAC-CDFT results. Comparing with the experimental values in Fig.~\ref{fig3-3}, the appearance of backbending is clearly seen for each band.

\subsection{Deformation evolution}

%%%%%%%%%%%%%%%%%%%%%%%%%%%%%%%%%%%%%%%%%%%%%%%%%%%%%%%%%%%%%%%%%%%%%%%%%%%%%%%%%%%%%%%%%%%%%%%%%%
\begin{figure*}
\centerline{
\includegraphics[scale=0.35,angle=0]{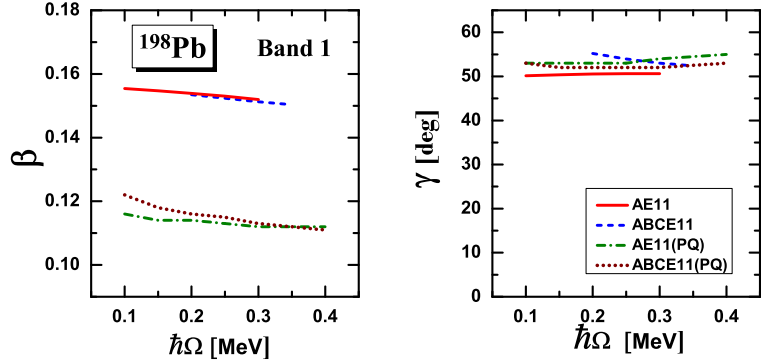}
} \caption{(Color online) Deformation parameters $\beta$ (left panels) and $\gamma$ (right panels) as  functions of the rotational frequency in the TAC-CDFT calculations
compared with the PQTAC results~\cite{Chmel2007Phys.Rev.C75} for band 1 in $^{198}\rm Pb$. Taken from Ref.~\cite{Yu2012Phys.Rev.C24318}.}
\label{fig3-4}
\end{figure*}
%%%%%%%%%%%%%%%%%%%%%%%%%%%%%%%%%%%%%%%%%%%%%%%%%%%%%%%%%%%%%%%%%%%%%%%%%%%%%%%%%%%%%%%%%%%%%%%%%%

The advantage of the TAC-CDFT calculations is that the nuclear shape and
deformation can be obtained self-consistently and automatically as outputs.
The deformation parameters $\beta$ and $\gamma$ as well as their evolutions for band 1 in $^{198}\rm Pb$ obtained in the TAC-CDFT calculations are compared with the PQTAC results~\cite{Chmel2007Phys.Rev.C75} in Fig.~\ref{fig3-4}. In the TAC-CDFT calculations,
the quadrupole deformations are around $\beta = 0.15$ and remain almost constant.  The PQTAC calculations produce the same tendency with slightly smaller deformations around $\beta =0.11$. Meanwhile, the deformation $\gamma$ vary between $47^\circ$ and $59^\circ$ which means small triaxiality  close to oblate axial symmetry in the TAC-CDFT calculations. This is consistent with the PQTAC results of Ref.~\cite{Chmel2007Phys.Rev.C75}.

%%%%%%%%%%%%%%%%%%%%%%%%%%%%%%%%%%%%%%%%%%%%%%%%%%%%%%%%%%%%%%%%%%%%%%%%%%%%%%%%%%%%%%%%%%%%%%%%%%
\begin{figure*}
\centerline{
\includegraphics[width=8cm]{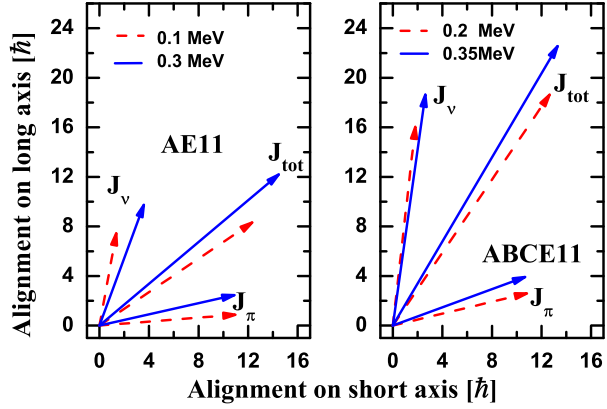}
} \caption{(Color online) Composition of the total angular momentum at both the minimum and the maximum
     rotational frequencies in TAC-CDFT calculations for band 1 in $^{198}\rm Pb$. Left (right) panel is result for the rotation before (after) backbending. Taken from Ref.~\cite{Yu2012Phys.Rev.C24318}.}
\label{fig3-5}
\end{figure*}
%%%%%%%%%%%%%%%%%%%%%%%%%%%%%%%%%%%%%%%%%%%%%%%%%%%%%%%%%%%%%%%%%%%%%%%%%%%%%%%%%%%%%%%%%%%%%%%%%%

\subsection{Shears mechanism}

One of the fundamental and important characteristic in magnetic rotation is the shears mechanism.
In Fig.~\ref{fig3-5}, the proton and neutron angular momentum vectors $\bm J_\pi$ and $\bm J_\nu$ as well as the total angular momentum vectors $\bm J_{\rm tot}=\bm J_\pi+\bm J_\nu$ at both the minimum and the maximum rotational frequencies in TAC-CDFT calculations for the band 1 in $^{198}\rm Pb$ are shown. The proton and neutron angular momenta $\bm J_\pi$ and $\bm J_\nu$ are defined as
\begin{equation}
    \bm J_\pi=\langle\bm{\hat J}_\pi\rangle=\sum_{p=1}^{Z}\langle p|\hat J|p\rangle, \quad\quad \bm J_\nu=\langle \bm{\hat J}_\nu\rangle=\sum_{n=1}^{N}\langle n|\hat J|n\rangle,
       \label{eq:AM}
\end{equation}
where the sum runs over all the proton (or neutron) levels occupied in the cranking wave function in the intrinsic system.

For the magnetic dipole bands in $^{198}\rm Pb$, the contributions to the angular momenta come mainly from the high $j$ orbitals, i.e., the $i_{13/2}$ neutron (s) as well as $h_{9/2}$ and $i_{13/2}$ protons. At the band head, the proton particles excited across the closed $Z = 82$ shell gap into the $h_{9/2}$ and $i_{13/2}$ orbitals contribute to the proton angular momentum along the short axis, and the neutron hole(s) at the upper end of the $i_{13/2}$ shell contribute to the neutron angular momentum along the long axis. By comparing the upper panels (before backbending) with the lower ones (after backbending) in Fig.~\ref{fig3-5}, one finds that after the backbending the neutron angular momentum vectors are considerably larger, because they contain the contributions of an aligned pair of $i_{13/2}$ neutron holes. Therefore, the proton and neutron angular momentum vectors form the two blades of the shears. With the frequency increasing, the two blades move toward each other and the closing of shears increase the angular momentum, while the direction of the total angular momentum stays nearly unchanged. In such a way, the shears mechanism is clearly observed.

\subsection{Electric and magnetic transitions}

%%%%%%%%%%%%%%%%%%%%%%%%%%%%%%%%%%%%%%%%%%%%%%%%%%%%%%%%%%%%%%%%%%%%%%%%%%%%%%%%%%%%%%%%%%%%%%%%%%
\begin{figure*}
\centerline{
\includegraphics[width=8cm]{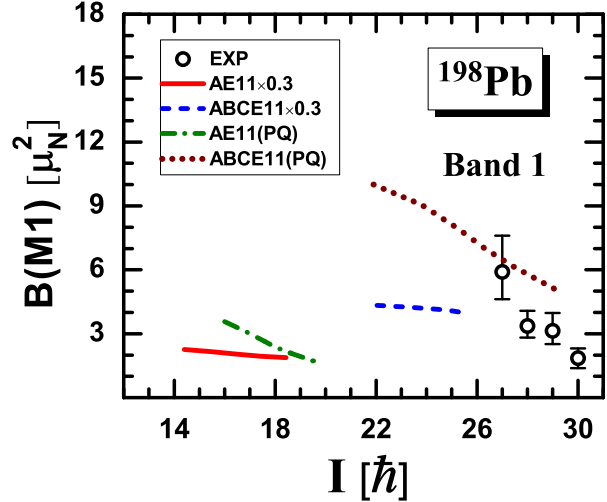}
}
\caption{(Color online) $B(M1)$ values as  functions of the total
angular momentum in the TAC-CDFT calculations compared
with the data and the PQTAC results~\cite{Chmel2007Phys.Rev.C75} for band 1 in $^{198}\rm Pb$. Circles and squares denote experimental data from DSAM~\cite{Clark1997Phys.Rev.Lett.1868} and RDM~\cite{Krucken1998Phys.Rev.C58}, respectively. Taken from Ref.~\cite{Yu2012Phys.Rev.C24318}.
}
\label{fig3-6}
\end{figure*}
%%%%%%%%%%%%%%%%%%%%%%%%%%%%%%%%%%%%%%%%%%%%%%%%%%%%%%%%%%%%%%%%%%%%%%%%%%%%%%%%%%%%%%%%%%%%%%%%%%

A typical characteristic of magnetic rotation is the strongly enhanced M1 transition probabilities which decrease with the spin.
In Fig.~\ref{fig3-6}, the calculated $B(M1)$ values as functions of the total angular momentum for the band 1 in $^{198}\rm Pb$
are shown in comparison with the data~\cite{Clark1997Phys.Rev.Lett.1868, Krucken1998Phys.Rev.C58} and the PQTAC results~\cite{Chmel2007Phys.Rev.C75}.
The TAC-CDFT calculations reproduce the decrease of the observed $B(M1)$ values with increasing spin. However, as observed already in earlier investigation~\cite{Madokoro2000Phys.Rev.C61301, Zhao2011Phys.Lett.B181}, the absolute values show discrepancies. As shown in Fig.~\ref{fig3-6}, one has to attenuate the results by a factor 0.3 in order to reproduce the absolute $B(M1)$ values. The same factor has been used in Refs.~\cite{Madokoro2000Phys.Rev.C61301, Zhao2011Phys.Lett.B181}. So far the origin of this attenuation factor is not understood in detail.
As discussed in Ref.~\cite{Zhao2011Phys.Lett.B181}, there are however several reasons: (a) Pairing correlations strongly affect the deformation and the levels in the neighborhood of the Fermi surface. This causes a strong reduction for the $B(M1)$ values with major contributions from the valence particles or holes. (b) The coupling to complex configurations such as particle vibrational coupling (Arima-Horie effect~\cite{Arima1954Prog.Theor.Phys.11,Arima2011Sci.China.Ser.G54}) leads in all cases to a quenching of the $B(M1)$ values for neutron configurations~\cite{Bauer1973Nucl.Phys.A209,Matsuzaki1988Prog.Theor.Phys.79}. (c) Meson exchange currents and higher corrections also cause a reduction of the effective $g$-factors for the neutrons~\cite{Towner1987Phys.Rep.155,Li2011Prog.Theor.Phys.125,Li2011Sci.China.Ser.G54}. However, it is not the absolute $B(M1)$ values, which characterize the shear bands, but rather the behavior of these values with increasing angular momentum.
On the other side, the absolute values of PQTAC results agree with the observed $B(M1)$ data and the attenuated TAC-CDFT results. However, they show a sharper decreasing trend as compared with the TAC-CDFT calculations. The agreement between the calculated and experimental $B(M1)$ values and their trend shows a convincing confirmation of the shears mechanism.

%%%%%%%%%%%%%%%%%%%%%%%%%%%%%%%%%%%%%%%%%%%%%%%%%%%%%%%%%%%%%%%%%%%%%%%%%%%%%%%%%%%%%%%%%%%%%%%%%%
\begin{figure*}
\centerline{
\includegraphics[width=8cm]{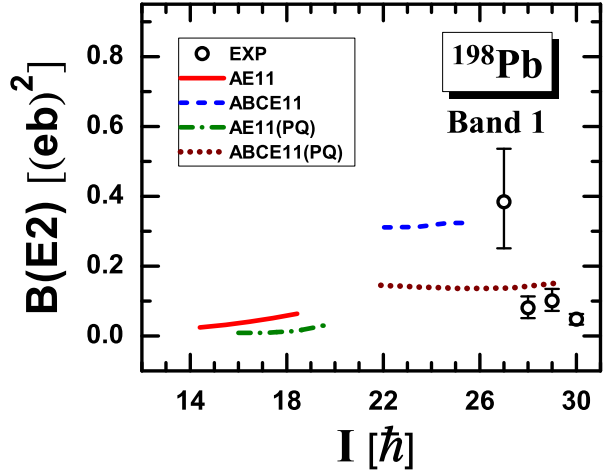}
}
\caption{(Color online) $B(E2)$ values as  functions of the total
angular momentum in the TAC-CDFT calculations compared
with the data from DSAM~\cite{Clark1997Phys.Rev.Lett.1868} and the PQTAC results~\cite{Chmel2007Phys.Rev.C75} for band 1. Taken from Ref.~\cite{Yu2012Phys.Rev.C24318}. }
\label{fig3-7}
\end{figure*}
%%%%%%%%%%%%%%%%%%%%%%%%%%%%%%%%%%%%%%%%%%%%%%%%%%%%%%%%%%%%%%%%%%%%%%%%%%%%%%%%%%%%%%%%%%%%%%%%%%

In contrast to the enhanced M1 transitions, the E2 transitions are weak for magnetic rotational bands.
In Fig.~\ref{fig3-7}, the calculated $B(E2)$ values as functions of the total angular momentum are shown and compared
with the DSAM-data of Ref.~\cite{Clark1997Phys.Rev.Lett.1868} and the PQTAC results of Ref.~\cite{Chmel2007Phys.Rev.C75} for band 1 in $^{198}\rm$Pb. The $B(E2)$ values in the TAC-CDFT calculations are in reasonable agreement
with the data and show a roughly constant trend. This is consistent with the nearly constant quadrupole deformation in each configuration calculated. Compared to the PQTAC results, the TAC-CDFT calculations predict larger $B(E2)$ values, in accordance with the larger deformations shown in Fig.~\ref{fig3-3}.

\section{Magnetic Rotation in other mass region}
%%%%%%%%%%%%%%%%%%%%%%%%%%%%%%%%%%%%
%%% beginning of Ni60
%%%%%%%%%%%%%%%%%%%%%%%%%%%%%%%%%%%

\subsection{$A\sim60$ mass region}

To date, the magnetic dipole bands observed
have been summarized in the nuclear chart in Fig.~\ref{fig1-3}.
The recent observations in $^{58}$Fe~\cite{Steppenbeck2012Phys.Rev.C85} and $^{60}$Ni~\cite{Torres2008Phys.Rev.C54318} is identified as the lightest mass region to exhibit magnetic rotation phenomenon and have extended the observed MR mass region to $A\sim60$ mass region.

In Ref.~\cite{Zhao2011Phys.Lett.B181},
TAC based CDFT calculation with PC-PK1~\cite{Zhao2010Phys.Rev.C54319} has been performed
for four magnetic dipole bands, denoted as M-1, M-2, M-3, and
M-4, reported in $^{60}\rm Ni$~\cite{Torres2008Phys.Rev.C54318}.
Same as in Ref.~\cite{Torres2008Phys.Rev.C54318},
the bands M-1 and M-4 are suggested to be built from the same type of
configurations, i.e.,
$\pi[(1f_{7/2})^{-1}(fp)^1]\otimes\nu[(1g_{9/2})^1(fp)^3]$.
For the bands M-2 and M-3, TAC based CDFT calculation clearly indicates that they
are respectively built from the configuration $\pi[(1f_{7/2})^{-1}(1g_{9/2})^1]\otimes\nu[(1g_{9/2})^1(fp)^3]$
and
$\pi[(1f_{7/2})^{-1}(fp)^1]\otimes\nu[(1g_{9/2})^2(fp)^2]$.
For simplicity, the above configurations
are referred as Config1, Config2, and Config3, respectively.

%%%%%%%%%%%%%%%%%%%%%%%%%%%%%%%%%%%%%%%%%%%%%%%%%%%%%%%%%%%%%%%%%%%%%%%%%%%%%%%%%%%%%%%%%%%%%%%%%%
\begin{figure*}[!htbp]
\centering
\includegraphics[width=12cm]{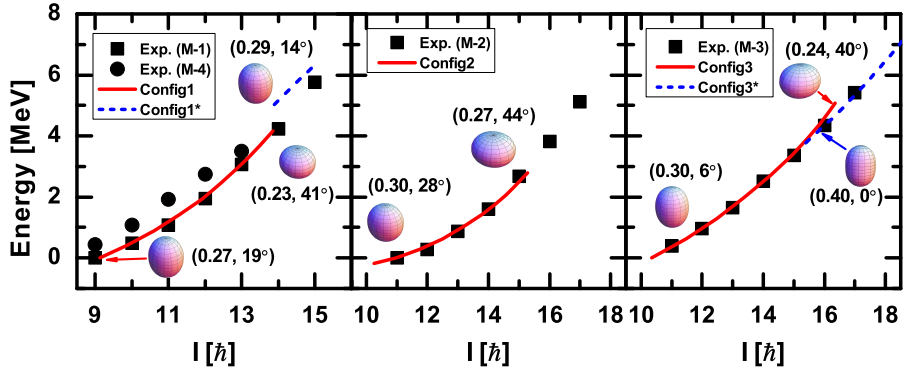}
\caption{(Color online) Energy spectra
obtained from the TAC-CDFT calculations in comparison with the available data for bands M-1 and
M-4 (left panel), M-2 (middle panel), as well as M-3 (right panel).
The energies at $I=9\hbar$, $I=11\hbar$, and $I=15\hbar$ are taken
as references in the left, middle and right panels, respectively. The evolutions of the nuclear shape $(\beta,\gamma)$ for bands M-1, M-2, and M-3 are also illustrated with the schematic pictures. Taken from Ref.~\cite{Zhao2011Phys.Lett.B181}.}
\label{fig4-1}
\end{figure*}
%%%%%%%%%%%%%%%%%%%%%%%%%%%%%%%%%%%%%%%%%%%%%%%%%%%%%%%%%%%%%%%%%%%%%%%%%%%%%%%%%%%%%%%%%%%%%%%%%%

The calculated energy spectra are shown in compared with the available data for the bands M-1 and M-4 (left
panel), M-2 (middle panel), as well as M-3 (right panel) for $^{60}\rm Ni$ in Fig.~\ref{fig4-1}.
In general, the experimental energies of the bands M-1, M-2, and M-3 are
reproduced very well by these TAC-CDFT calculations.
However, the assigned configuration for each of these bands could not be followed in
the calculations up to the highest spin observed, i.e., convergent
results could be obtained only up to $\sim14\hbar$ for Config1,
$\sim15\hbar$ for Config2, and $\sim16\hbar$ for Config3. These are connected with
the configuration change and shape evolution~\cite{Zhao2011Phys.Lett.B181}.

With the increase of the rotational frequency, the configurations, $\pi[(1f_{7/2})^{-1}(fp)^1]\otimes\nu[(1g_{9/2})^{1}(fp)^4(1f_{7/2})^{-1}]$ (Config1*)
and
$\pi[(1f_{7/2})^{-2}(fp)^2]\otimes\nu[(1g_{9/2})^{2}(fp)^3(1f_{7/2})^{-1}]$ (Config3*)
will strongly compete with Config1 and Config3, respectively.
In other words, one observes
a neutron pair broken in the $f_{7/2}$ shell at $I=15\hbar$ in band M-1,
and
the excitation of a unpaired proton from the $f_{7/2}$ shell to the $fp$ orbital
and a neutron pair broken in the $f_{7/2}$ shell
at $I=16\hbar$ in the band M-3.

The shape evolutions of bands M-1, M-2, and M-3 are also shown in Fig.~\ref{fig4-1}.
It is interesting to note that the nucleus changes its shape from prolate-like to oblate-like with the frequency in Config1, Config2, and Config3, and comes back to a prolate-like deformation with the configuration changing from Config1 to Config1*, and Config3 to Config3*. In particular, the nucleus with Config3* has a relatively large deformation ($\beta\sim0.4$) with axial symmetry.

%%%

%%%%%%%%%%%%%%%%%%%%%%%%%%%%%%%%%%%%%%%%%%%%%%%%%%%%%%%%%%%%%%%%%%%%%%%%%%%%%%%%%%%%%%%%%%%%%%%%%%
\begin{figure}[!htbp]
\centering
\includegraphics[width=8cm]{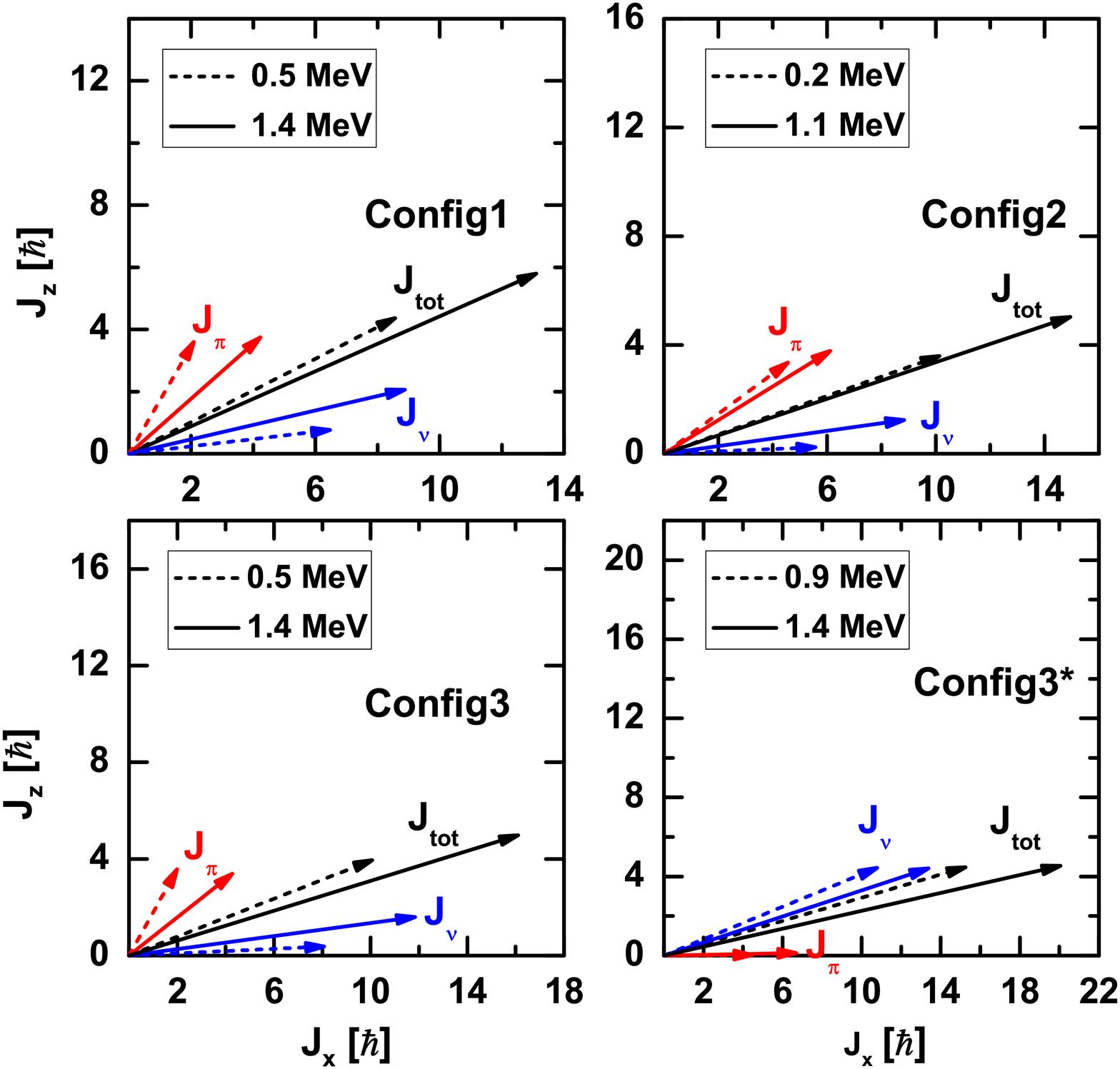}
\caption{(Color online) Composition of the total angular momentum at both the bandhead and the maximum rotational frequency in the TAC-CDFT calculations with the configurations of Config1, Config2, Config3, and Config3*. Taken from Ref.~\cite{Zhao2011Phys.Lett.B181}.}
\label{fig4-2}
\end{figure}
%%%%%%%%%%%%%%%%%%%%%%%%%%%%%%%%%%%%%%%%%%%%%%%%%%%%%%%%%%%%%%%%%%%%%%%%%%%%%%%%%%%%%%%%%%%%%%%%%%

For the magnetic rotation bands in $^{60}\rm Ni$, the proton and neutron angular momentum vectors $\bm{J}_\pi$ and $\bm{J}_\nu$ defined in Eq.~(\ref{eq:AM}) as well as the total angular momentum vector $\bm{J}_{\rm tot}$ at both the bandhead and the maximum rotational frequency in the TAC-CDFT calculations with the configurations of Config1, Config2, Config3, and Config3*, are shown in Fig.~\ref{fig4-2}.

For the bands built on Config1, Config2, and Config3, the contributions to the angular momenta come mainly from the high $j$ orbitals, i.e., the $g_{9/2}$ neutron(s) and the $f_{7/2}$ proton. At the bandhead, the neutron particle(s) filling the bottom of the $g_{9/2}$ shell mainly contribute to the neutron angular momentum along the
$x$-axis, and the proton hole at the upper end of the $f_{7/2}$ shell mainly contributes to the proton angular momentum along the $z$-axis. They form the two blades of the shears. As the frequency increases, the two blades move toward each other to provide larger angular momentum, while the direction of the total angular momentum stays nearly unchanged. In this way, the shears mechanism is clearly seen.

One should notice that the proton particle in the $g_{9/2}$ orbital also give substantial contributions to the proton angular momentum in the case of Config2. As a result, $\bm{J}_\pi$ has not only a large $J_z$ component but also a substantial $J_x$ component even at the bandhead. Thus, the shears angle $\Theta$, the angle between $\bm{J}_\pi$ and $\bm{J}_\nu$, is not as large as those of Config1 and Config3, and decreases only by a small amount with increasing rotational frequency.

For the Config3*, as the two proton holes in the $f_{7/2}$ orbital are paired, the proton angular momentum comes mainly from the particles in the $fp$ shell, which aligns along the $x$-axis. The neutron hole in the $f_{7/2}$ orbital gives substantial contributions to the neutron angular momentum, which leads to a large $J_z$ component. Higher spin states in the band are created by aligning the neutron angular momentum towards the $x$-axis.
Considering the large axially symmetric prolate deformation as shown in Fig.~\ref{fig4-1}, the mechanism of producing higher spin states with Config3* is electric rotation rather than magnetic rotation. Therefore, a transition from magnetic rotation to electric rotation is observed in Config3*.

Recently, the high spin structure in $^{58}$Fe has been investigated by
heavy-ion induced fusion-evaporation reactions
at Gammasphere~\cite{Steppenbeck2012Phys.Rev.C85}.
The magnetic rotational bands observed have been interpreted with the TAC-CDFT,
which concludes that $^{58}$Fe is the
lightest nucleus exhibiting magnetic rotation.

%%%%%%%%%%%%%%%%%%%%%%%%%%%%%%%%%%%%
%%% end of Ni60
%%%%%%%%%%%%%%%%%%%%%%%%%%%%%%%%%%%

%%%%%%%%%%%%%%%%%%%%%%%%%%%%%%%%%%%%
%%% beginning of Gd142
%%%%%%%%%%%%%%%%%%%%%%%%%%%%%%%%%%%

\subsection{$A\sim140$ mass region}

For $A\sim140$ mass region,
$\pi h_{11/2}$ particles combined with $\nu h_{11/2}$ holes
satisfy the high-$j$ configurations for magnetic rotation.
In fact, the magnetic rotation in $A\sim140$ mass region have been identified in Tb, Gd, Eu, Sm, Pm, Nd, Pr, Ce, La, Ba, Cs, Xe, Te and Dy isotopes~(see \cite{Amita2000At.DataNucl.DataTables283} and the references therein).

For $^{142}$Gd, five bands denoted as DB1, DB2, DB3,
DB4 and DB5, have been observed and four of them have been
interpreted as magnetic rotation bands with the configurations $\pi
h_{11/2}^{2}\otimes\nu h_{11/2}^{-2}$, $\pi h_{11/2}^{2}\otimes\nu
h_{11/2}^{-4}$, $\pi h_{11/2}^{1}\otimes\pi g_{7/2}^{-1}\nu h_{11/2}^{-2}$,
and $\pi h_{11/2}^{1}\otimes\pi g_{7/2}^{-1}\nu h_{11/2}^{-4}$~\cite{Lieder2002Euro.Phys.J.A13,Pasternak2005Euro.Phys.J.A23},
respectively.
In Ref.~\cite{Olbratowski2002APPB.33.389}, the shears mechanism and
the spectrum of the band DB1 in $^{142}$Gd has been investigated by
the tilted axis cranking Skyrme Hartree-Fock method.
In Ref.~\cite{Peng2008Phys.Rev.C24313},
 TAC-CDFT with the meson-exchange interaction PK1~\cite{Long2004Phys.Rev.C34319}
 has been applied for the magnetic rotational band DB1 in
$^{142}$Gd based on the configuration $\pi
h^{2}_{11/2} \otimes\nu h^{-2}_{11/2}$.
For simplicity, the configurations $\pi h_{11/2}%
^{2}\otimes\nu h_{11/2}^{-2}$ and $\pi\lbrack h_{11/2}^{2}g_{7/2}^{-1}%
d_{5/2}^{1}]\otimes\nu h_{11/2}^{-2}$ are refereed as Config1
and Config1*, respectively.

%%%%%%%%%%%%%%%%%%%%%%%%%%%%%%%%%%%%%%%%%%%%%%%%%%%%%%%%%%%%%%%%%%%%%%%%%%%%
{\normalsize \begin{figure}[tbh]
{\normalsize \centering
\includegraphics[height=7cm]{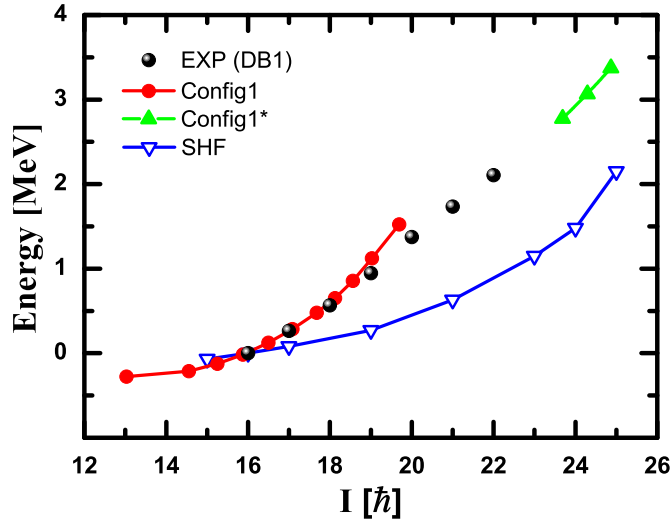}  }\caption{(Color online)
The energy as a function of the total spin in TAC-CDFT calculation~\cite{Peng2008Phys.Rev.C24313} for
Config1 (full red dots) and Config1*
(green triangles up) in comparison with the data for the band DB1 in
$^{142}$Gd (filled back circles)~\cite{Pasternak2005Euro.Phys.J.A23}. The non-relativistic
SHF result (blue triangles down )~\cite{Olbratowski2002APPB.33.389} is also included. }
\label{fig4-3}%
\end{figure}}
%%%%%%%%%%%%%%%%%%%%%%%%%%%%%%%%%%%%%%%%%%%%%%%%%%%%%%%%%%%%%%%%%%%%%%%%%%%%

The energies as functions of the total
angular momentum for the Config1 calculation (full red dots) and the Config1*
calculation (green triangles ups) for the magnetic dipole band DB1 in $^{142}$Gd are shown in Fig.~\ref{fig4-3}. They are
compared with the available data (filled circles)~\cite{Pasternak2005Euro.Phys.J.A23}
and the non-relativistic SHF results from Ref.~\cite{Olbratowski2002APPB.33.389} (triangles
down). As no link to the ground state is observed, the recommended band head spin $I=16$ $\hbar$
of Ref.~\cite{Pasternak2005Euro.Phys.J.A23} is adopted, and the energy at $I=16$ $\hbar$ are taken
as a reference for both the RMF values and the nonrelativistic SHF calculation. In general, the energies in RMF calculations achieve
better agreement with the data as compared with the SHF results.

%%%%%%%%%%%%%%%%%%%%%%%%%%%%%%%%%%%%%%%%%%%%%%%%%%%%%%%%%%%%%%%%%%%%%%%%%%%%
{\normalsize \begin{figure}[tbh]
{\normalsize \centering
\includegraphics[height=7cm]{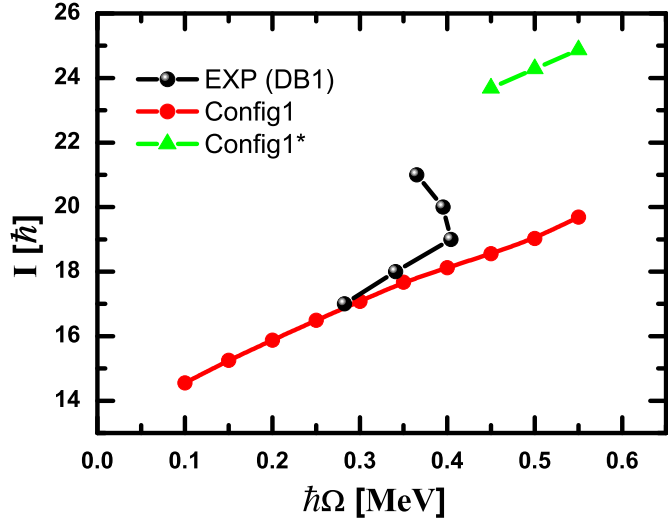}  }\caption{(Color online)
The total angular momentum as a function of the rotational frequency in
 TAC-CDFT calculation~\cite{Peng2008Phys.Rev.C24313} for the configurations $\pi h^{2}%
_{11/2}\otimes\nu h^{-2}_{11/2}$ (Config1, open circles) and $\pi[h^{2}%
_{11/2}g^{-1}_{7/2}d^{1}_{5/2}] \otimes\nu h^{-2}_{11/2}$ (Config1*, triangles
up) in comparison with the data for DB1 in $^{142}$Gd (filled
circles)~\cite{Pasternak2005Euro.Phys.J.A23}. }%
\label{fig4-4}%
\end{figure}}
%%%%%%%%%%%%%%%%%%%%%%%%%%%%%%%%%%%%%%%%%%%%%%%%%%%%%%%%%%%%%%%%%%%%%%%%%%%%

The total angular momenta are shown as functions
of the rotational frequency and results of Config1 (full red dots) and Config1*
(green triangles up) calculations are compared with data (filled circles) of
Ref.~\cite{Pasternak2005Euro.Phys.J.A23} in Fig.~\ref{fig4-4}.
The total angular momenta in Config1 linearly increase with the rotational
frequency and agree with the data till $\hslash\Omega=0.40$ MeV. After
$\hslash\Omega>0.40$ MeV, a up-bending is observed for the data. This
up-bending can not be reproduced by the smooth behavior in either Config1 or
Config1* calculations.

%%%%%%%%%%%%%%%%%%%%%%%%%%%%%%%%%%%%
%%% en of Gd142
%%%%%%%%%%%%%%%%%%%%%%%%%%%%%%%%%%%
%%%%%%%%%%%%%%%%%%%%%%%%%%%%%%%%%%%%%

\subsection{$A\sim 80$ and $A\sim110$ mass regions}

For $A\sim80$ mass region,
$\pi g_{9/2}$ particles combined with $\nu g_{9/2}$ holes
satisfy the high-$j$ configurations for magnetic rotation.
In fact, the magnetic rotation in $A\sim80$
mass region have been identified in Rb, Kr and Br
isotopes~(see \cite{Amita2000At.DataNucl.DataTables283} and the references therein).

As the first relativistic investigation of the magnetic rotation,
three-dimensional cranking CDFT has been developed and applied for
$^{84}$Rb~\cite{Madokoro2000Phys.Rev.C61301}.
The proton configuration is fixed
to be $\pi(pf)^{7}(1g_{9/2})^{2}$ with respect to the $Z=28$ magic number
and $\nu(1g_{9/2})^{-3}$ with respect to the $N=50$ magic number is adopted for the neutron configuration.
The signals of the shears mechanism, such as the nearly constant tilt angle and
the smooth decreases of the shears angle and of
the $B(M1)/B(E2)$ ratio, are well reproduced. The detailed discussion
can be seen in Ref.~\cite{Madokoro2000Phys.Rev.C61301}.
Because of the numerical complexity, so far,
the three-dimensional cranking CDFT has
been applied only for the magnetic rotation in $^{84}$Rb~\cite{Madokoro2000Phys.Rev.C61301}.

For $A\sim 110$ mass region,
$\pi h_{11/2}$ particles combined with $\nu g_{9/2}$ holes
satisfy the high-$j$ configurations for magnetic rotation.
In fact, the magnetic rotation in $A\sim110$
mass region have been identified in Cd, In, Sn, Sb, and Te
isotopes~(see \cite{Amita2000At.DataNucl.DataTables283} and the references therein).
Using the recently developed TAC-CDFT based on point-coupling interactions,
the magnetic rotation bands in $^{113,114}$In in $A\sim110$ mass region
are well reproduced successfully~\cite{Ma2012Eur.Phys.J.A48,Li2012Nucl.Phys.A892}.

In order to explore the MR in the lightest mass region,
the TAC-CDFT calculation has been performed for $^{22}$F with the configuration $\pi d_{5/2}\otimes\nu
d^{-1}_{5/2}$ in Ref.~\cite{Peng2010Chin.Phys.Lett.27}.
The possible existence of magnetic rotation is
suggested for $^{22}$F via investigating the spectra, the
relation between the rotational frequency and the angular momentum,
the electromagnetic transition probabilities $B(M1)$ and $B(E2)$
together with the shears mechanism characteristic of magnetic
rotation.

\section{Antimagnetic rotation}

%%%%%%%%%%%%%%%%%%%%%%%%%%%%%%%%%%%%
%%% beginning of Cd105
%%%%%%%%%%%%%%%%%%%%%%%%%%%%%%%%%%%
\subsection{Concept}

As mentioned in the introduction,
 in analogy with an antiferromagnet, antimagnetic rotation" (AMR) [9] is predicted to
occur in some specific nearly spherical nuclei, in which the subsystems of valence protons
(neutrons) are aligned back to back in opposite directions and nearly perpendicular to the
orientation of the total spin of the valence neutrons (protons). Such arrangement of the
proton and neutron angular momenta also breaks the rotational symmetry in these nearly
spherical nuclei and causes excitations with rotational character on top of this bandhead as ¡°antimagnetic rotation¡±~\cite{Frauendorf1996272,Frauendorf2001Rev.Mod.Phys.463}.

To date, antimagnetic rotations have been reported in Cd isotopes including
$^{105}\rm Cd$~\cite{Choudhury2010Phys.Rev.C61308}, $^{106}\rm Cd$~\cite{Simons2003Phys.Rev.Lett.162501}, $^{108}\rm Cd$~\cite{Simons2005Phys.Rev.C24318,Datta2005Phys.Rev.C41305}, and $^{110}\rm Cd$~\cite{Roy2011Phys.Lett.B322}. The other candidates include $^{109}\rm Cd$~\cite{Chiara2000Phys.Rev.C34318}, $^{100}\rm Pd$~\cite{Zhu2001Phys.Rev.C41302}, and $^{144}\rm Dy$~\cite{Sugawara2009Phys.Rev.C64321}.
In order to apply the CDFT for antimagnetic rotation phenomenon,
the newly developed TAC-CDFT based on point-coupling interactions are used to investigate
antimagnetic rotation (AMR) in $^{105}\rm Cd$ in a fully self-consistent and microscopic way in Ref.~\cite{Zhao2011Phys.Rev.Lett.122501}.

\subsection{Energy spectrum}

%%%%%%%%%%%%%%%%%%%%%%%%%%%%%%%%%%%%%%%%%%%%%%%%%%%%%%%%%%%%%%%%%%%%%%%%%%%%
\begin{figure}[!htbp]
\centering
 \includegraphics[width=8cm]{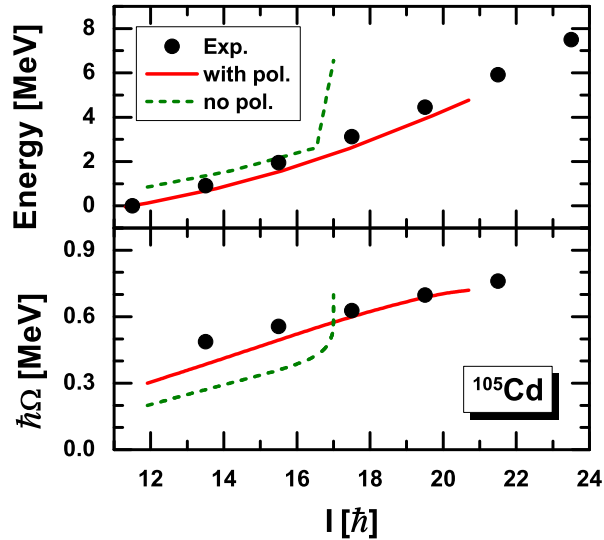}
\caption{(color online) Energy (upper panel) and rotational frequency (lower panel) as functions of the total angular momentum. The fully self-consistent solution (solid lines) and that neglecting polarization (dashed lines) are compared with the data~\cite{Choudhury2010Phys.Rev.C61308} (solid dots). The energy at $I = 23/2\hbar$ is taken as reference in the upper panel.  Taken from Ref.~\cite{Zhao2011Phys.Rev.Lett.122501}.}
\label{fig5-1}
\end{figure}
%%%%%%%%%%%%%%%%%%%%%%%%%%%%%%%%%%%%%%%%%%%%%%%%%%%%%%%%%%%%%%%%%%%%%%%%%%%%

For the TAC-CDFT calculation~\cite{Zhao2011Phys.Rev.Lett.122501} based on point-coupling interactions PC-PK1~\cite{Zhao2010Phys.Rev.C54319}
for AMR band in $^{105}\rm Cd$, the odd neutron occupies the lowest level in the $h_{11/2}$ shell
 and the remaining nucleons are treated self-consistently
 by filling the orbitals according to their energy from the bottom of the well.
 This automatically leads to the configuration for AMR.

In Fig.~\ref{fig5-1},  the TAC-CDFT calculated energy and the rotational frequency (solid lines)~\cite{Zhao2011Phys.Rev.Lett.122501}
are compared with data~\cite{Choudhury2010Phys.Rev.C61308}. In the upper panel it can be clearly seen that, apart from the bandhead, the experimental energies are reproduced excellently by the present self-consistent calculations. In the lower panel it is found that the calculated total angular momenta agree well with the data and increase almost linearly with increasing frequency. This indicates that the  moment of inertia is nearly constant and well reproduced by the present calculations.

\subsection{Two shearslike mechanism}

%%%%%%%%%%%%%%%%%%%%%%%%%%%%%%%%%%%%%%%%%%%%%%%%%%%%%%%%%%%%%%%%%%%%%%%%%%%%
\begin{figure}[!htbp]
\centering
 \includegraphics[width=8cm]{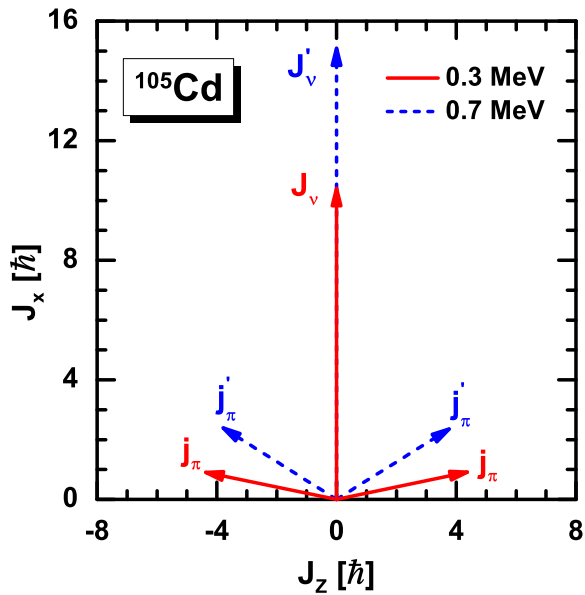}
\caption{(color online) Angular momentum vectors of neutrons $\bm{J}_\nu$ and the two $g_{9/2}$ proton holes $\bm{j}_\pi$ at both the bandhead ($\hbar\Omega=0.3$~MeV)
and the maximum rotational frequency.  Taken from Ref.~\cite{Zhao2011Phys.Rev.Lett.122501}.}
\label{fig5-2}
\end{figure}

%%%%%%%%%%%%%%%%%%%%%%%%%%%%%%%%%%%%%%%%%%%%%%%%%%%%%%%%%%%%%%%%%%%%%%%%%%%%

In order to demonstrate the two shearslike mechanism in $^{105}\rm Cd$ in Fig.~\ref{fig5-2}, there was shown that both at the bandhead and at the maximum rotational frequency the angular momentum vectors of the two $g_{9/2}$ proton-holes $\bm{j}_\pi$ and of the neutrons $\bm{J}_\nu=\sum_{n}\bm{j}^{(n)}_\nu$
where $n$ runs over all the occupied neutron levels. At the bandhead, the two proton angular momentum vectors $\bm{j}_\pi$ are pointing opposite to each other and are nearly perpendicular to the vector $\bm{J}_\nu$. They form the blades of the two shears. With increasing $\Omega$ the gradual alignment of the vectors $\bm{j}_\pi$ of the two $g_{9/2}$ proton holes toward the vector $\bm{J}_\nu$ generates angular momentum while the direction of the total angular momentum stays unchanged. This leads to the closing of the two shears. The two shearslike mechanism can thus be clearly seen, and it is consistent with the previous works~\cite{Frauendorf2001Rev.Mod.Phys.463,Simons2003Phys.Rev.Lett.162501}.

%%%%%%%%%%%%%%%%%%%%%%%%%%%%%%%%%%%%%%%%%%%%%%%%%%%%%%%%%%%%%%%%%%%%%%%%%%%%
\begin{figure}[!htbp]
\centering
 \includegraphics[width=8cm]{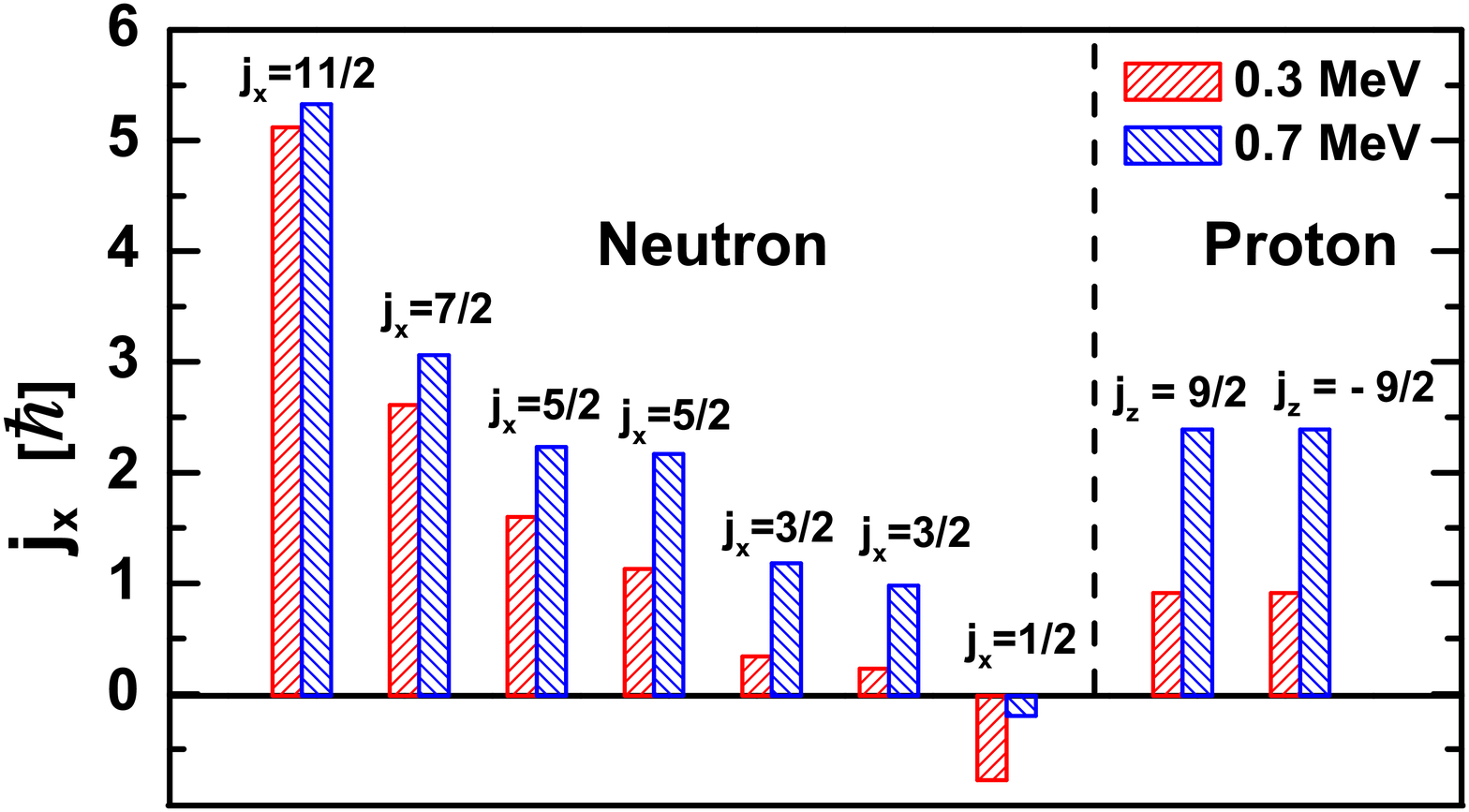}
\caption{(color online) Alignment of the valence neutrons (left side) and proton holes (right side) at both the bandhead ($\hbar\Omega=0.3$~MeV)
and the maximum rotational frequency.  Taken from Ref.~\cite{Zhao2011Phys.Rev.Lett.122501}.}
\label{fig5-3}
\end{figure}
%%%%%%%%%%%%%%%%%%%%%%%%%%%%%%%%%%%%%%%%%%%%%%%%%%%%%%%%%%%%%%%%%%%%%%%%%%%%

In a microscopic calculation, there is no inert core and all the energy and angular momentum comes from the particles.
The contributions of the valence neutrons and proton holes to the angular momentum $J_x$ at both the bandhead and the maximum rotational frequency are shown
in Fig.~\ref{fig5-3}. It is found that the contributions come mainly from high-$j$ orbitals, i.e., from $g_{9/2}$ proton holes as well as from $h_{11/2}$ and $g_{7/2}$ neutrons. In order to provide a simple picture which can be compared with the core angular momentum given in
Ref.~\cite{Frauendorf2001Rev.Mod.Phys.463}, one can estimate the ``corelike'' angular momentum in the present framework by excluding the contributions of three valence neutrons, shown in the left three columns in Fig.2, from the total neutron angular momentum. It is found that the ``core'' contributes about 3 $\hbar$ when the frequency $\Omega$ increases from the bandhead to the maximum value.

For the protons, only the two holes in the $g_{9/2}$ shell contribute. As shown in Fig.~\ref{fig5-2}, they cancel each other in the $z$ direction giving non-negligible contributions to the angular momentum along $x$ axis even at the bandhead. With growing frequency, the proton angular momentum in $x$ direction increases because of the alignment of the two proton hole blades. For the neutrons, on the other hand, we have only contributions above the $N=50$ shell. One neutron sits in the $h_{11/2}$ orbit and the other six are, because of considerable mixing, distributed over the $g_{7/2}$ and $d_{5/2}$ orbitals.  As $\Omega$ grows, the contributions of the aligned orbitals with $j_x=11/2$ and $9/2$ do not change much and the increase in angular momentum is generated mostly by the alignment of orbitals with low $j_x$ values. This microscopic calculation shows that the interpretation given in Ref.~\cite{Choudhury2010Phys.Rev.C61308} is only partially justified: we clearly have two proton holes in the $g_{9/2}$ and one neutron particle in the $h_{11/2}$ orbit, but, due to the mixing of orbits with lower $j$ values the other neutrons are distributed over several subshells above the $N=50$ core and the increasing angular momentum results from the alignment of the proton holes and the mixing within the neutron orbitals. Because of this strong mixing between the neutrons, a core needed for the phenomenological model in Ref.~\cite{Clark2000Annu.Rev.Nucl.Part.Sci.1} cannot really be defined.

\subsection{Electric transition probability and deformation}

%%%%%%%%%%%%%%%%%%%%%%%%%%%%%%%%%%%%%%%%%%%%%%%%%%%%%%%%%%%%%%%%%%%%%%%%%%%%
\begin{figure}[!htbp]
\centering
 \includegraphics[width=8cm]{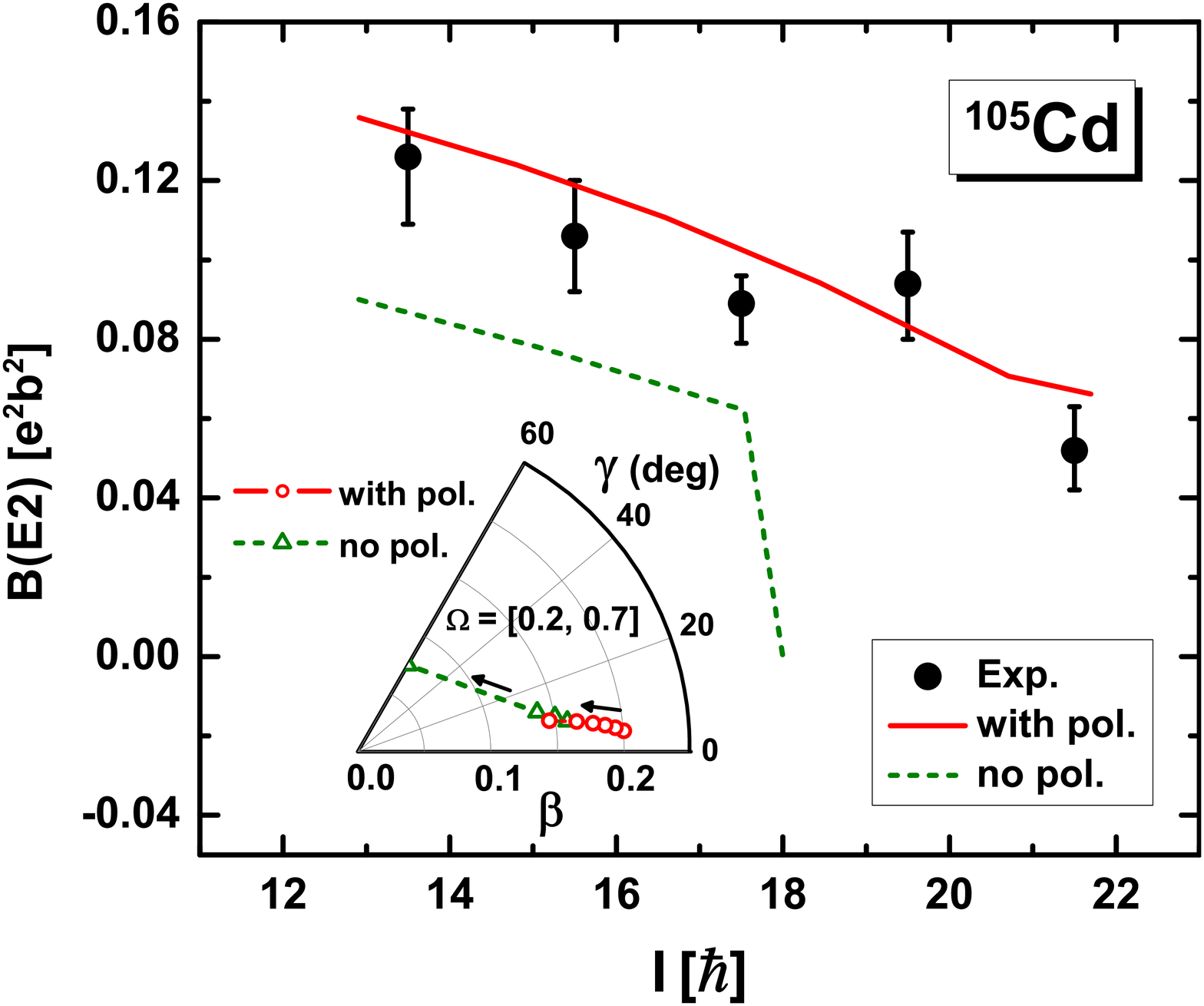}
\caption{(color online) $B(E2)$ values as a function of the angular momentum. Solutions with (solid line) and without (dashed line) polarization are compared with data~\cite{Choudhury2010Phys.Rev.C61308} (solid dots). Inset: Deformations $\beta$ and $\gamma$ driven by the increasing rotational frequency whose direction is indicated by arrows.  Taken from Ref.~\cite{Zhao2011Phys.Rev.Lett.122501}}
\label{fig5-4}
\end{figure}
%%%%%%%%%%%%%%%%%%%%%%%%%%%%%%%%%%%%%%%%%%%%%%%%%%%%%%%%%%%%%%%%%%%%%%%%%%%%

AMR is characterized by weak E2 transitions decreasing with increasing spin. In Fig.~\ref{fig5-4}, the calculated $B(E2)$ values (solid lines) are compared with the available data~\cite{Choudhury2010Phys.Rev.C61308}. It is found that the resulting $B(E2)$ values are very small ($< 0.14~e^2b^2$) and in very good agreement with the data. Furthermore, the fact that the $B(E2)$ values decrease with spin is in agreement with the interpretation by the two shearslike mechanism.

The decrease of the $B(E2)$ values can be understood by the changes in the nuclear deformation. As shown in the inset of Fig.~\ref{fig5-4}, with increasing frequency, the nucleus undergoes a rapid decrease of $\beta$ deformation from $0.2$ to $0.14$ at a small and near-constant triaxiality ($\gamma\leq 9^\circ$). As usual, it is found that the deformation of the charge distribution, responsible for the $B(E2)$ values, changes in a similar manner. Therefore, one can conclude that the alignment of the proton and neutron angular momenta, i.e., the two shearslike mechanism, is accompanied by a transition from prolate towards nearly spherical shape.

\subsection{Core polarization}

In order to investigate the importance of the polarization effects induced by the two proton holes, which is taken into account fully in the TAC-CDFT calculation, an additional calculation without polarization has been carried out in Ref~\cite{Zhao2011Phys.Rev.Lett.122501}. For this purpose, at each frequency the $^{107}\rm Sn$ core is firstly calculated in Ref.~\cite{Zhao2011Phys.Rev.Lett.122501}, where the two proton holes are filled. This results in a filled nearly spherical $g_{9/2}$ shell. In the next step, the self-consistency (dashed lines in Figs.~\ref{fig5-1} and \ref{fig5-4}) is neglected and the band in $^{105}\rm Cd$
is calculated, in which two protons are removed from the $g_{9/2}$ shell using, at each frequency, the corresponding nearly spherical potentials $S$ and $V^\mu$ obtained in the calculations of the $^{107}\rm Sn$ core. As shown in the upper panel of Fig.~\ref{fig5-1}, the energy is reduced only slightly by polarization in the lower part of the spectrum. At the same time the relation between angular velocity and angular momentum is considerably changed in the lower part of Fig.~\ref{fig5-1}. Without polarization, a much smaller frequency $\Omega$ is necessary to reach the same angular momentum as with polarization. In addition, without polarization, there is a maximal angular momentum of roughly 17 $\hbar$. Higher values cannot be reached even at rather high frequencies.

This behavior can be well understood from the evolution of the deformation parameters shown in the inset of Fig.~\ref{fig5-4}. Without polarization we use at each frequency the potentials of the core-nucleus $^{107}\rm Sn$, where the deformation is relatively small. Angular momentum can only be produced by alignment of neutron particles along the rotational axis leading at $\hbar\Omega=0.5$ MeV to an oblate shape with a rotation around the symmetry axis. Removing two protons would lead, if polarization is taken into account, to a larger prolate configuration with lower energy and hindering alignment. Therefore, in the lower part of the band, it is easier to produce angular momentum without polarization, where the deformation is small. On the other side the oblate deformation keeps the high-$j$ proton holes in the $j_x=\pm 1/2$ orbitals of the $g_{9/2}$ shell pairwise occupied and hinders their alignment. Above $\hbar\Omega=0.5$ MeV we reach the maximum angular momentum of the neutron configuration. With polarization, because of the prolate deformation, we have much more mixing and therefore can reach larger angular momentum by aligning the protons. This can also be seen in Fig.~\ref{fig5-4} that the E2 transitions do survive with polarization when $\hbar\Omega\ge0.5$ MeV. Therefore, it is of importance to emphasize that polarization effects play a very important role in the self-consistent microscopic description of AMR bands.
%%%%%%%%%%%%%%%%%%%%%%%%%%%%%%%%%%%%
%%% en of Cd105
%%%%%%%%%%%%%%%%%%%%%%%%%%%%%%%%%%%

\section{Summary and perspectives}

In the past decades, the rotational-like sequences in near-spherical or weakly deformed nuclei
have attracted significant attentions. This phenomenon, known as magnetic rotation, has
been extensively explored experimentally and theoretically.
With its many success in describing nuclear phenomena in
stable as well as in exotic nuclei, the CDFT has been generalized to the cranking CDFT, the tilted axis cranking CDFT and three-dimensional cranking CDFT and applied for electric and magnetic rotations all over the nuclear chart.
In particular, the newly developed TAC-CDFT based on point-coupling interactions
includes significant improvements and reduces computation time, which makes it possible to perform systematic investigation.

This review provides an overview of the experiential and theoretical status
of MR and AMR, a sketch of the CDFT as well as TAC-CDFT based on point-coupling interactions, and
followed by the summary of the TAC-CDFT
descriptions for the MR and AMR bands.

The shears bands in the nuclei $^{198}\rm Pb$ is the most well-known example of MR.
The TAC-CDFT calculated energy spectra, the relation between spin and rotational frequency,
the deformation parameters and reduced M1 and E2 transition probabilities are discussed and
compared with data. By choosing $^{60}\rm Ni$, $^{84}$Rb, and $^{142}$Gd as examples, the success of the TAC-CDFT
in describing MR for $A\sim 60$,  $A\sim 80$, and $A\sim 140$ mass regions has been demonstrated.
By reproducing the experimental energy spectrum and the $B(E2)$ values combining with the examination of the
core polarization and deformation evolution, an antimagnetic rotation band in $^{105}$Cd has been investigated in a fully self-consistent and microscopic way by the TAC-CDFT. The two shearslike mechanism in AMR has been clearly
illustrated.

One should note that the pairing correlations have been neglected in the calculations presented in this review. In fact, the pairing correlations are negligible for many MR and AMR bands in which the high $j$ oribitals near the Fermi surface are blocked and thus reduce the pairing effects. However, in some specific MR and AMR bands, the low $j$ orbitals also play important role in the rotation excitations. Therefore, it is very important to investigate the pairing effects on the MR and AMR in the framework of the tilted axis cranking covariant density functional theory. Due to the time-reversal symmetry broken, the simple BCS method is not valid and one should resort to the Bogoliubov method. In such case, the treatment of particle projection is usually necessary since the particle number conservation is violated in the Bogoliubov transformation.

There is another way known as Particle Number Conservation (PNC) method~\cite{Zeng19831,Zeng1994Phys.Rev.C1388}, which can treat the pairing correlation exactly and has been implemented successfully in the cranking Nilsson model~\cite{Zeng1994Phys.Rev.C1388,Wu2011Phys.Rev.C34323,Zhang2012Phys.Rev.C14324}. Similar method, the Shell-model-like Approach (SLAP), has also been applied to treat the pairing correlations in the framework of CDFT~\cite{MENG2006Front.Phys.China38}. The SLAP has the advantage that the paring correlation is treated exactly and thus the particle number conservation is not violated in the calculations. Therefore, it would be very interesting to introduce this method to the present tilted axis cranking covariant density functional theory to investigate the pairing effects in the tilted axis cranking calculations.

During the past decades, the three-dimensional cranking calculation has attracted wide attentions since it is connected with the novel phenomenon of chiral rotation in nuclei. Therefore, it is also a very interesting to extend the tilted axis cranking covariant density functional theory to three-dimensional cranking case with the point-coupling interaction. Similar work has been performed with the meson-exchange interactions. However, efforts should be made to simplify its numerical complicity  to investigate the chiral rotation.

\begin{acknowledgments}
We would like to express our gratitude to all the friends and collaborators, who contributed to the investigations presented here, in particular to S. Frauendorf, H. Z. Liang, H. Madokoro, M. Matsuzaki, P. Ring, S. Yamaji, and L. F. Yu. This work was supported in part by the Major State 973 Program of China (Grant No. 2013CB834400), the Natural Science Foundation of China (Grants No. 10975007, No. 10975008, No. 11175002, No. 11105005), the Research Fund for the Doctoral Program of Higher Education under Grant No. 20110001110087, and the China Postdoctoral Science Foundation Grant No. 2012M520101.
\end{acknowledgments}

\newpage %Just because of unusual number of tables stacked at end

\end{document}